\documentclass[letterpaper,11pt]{article}
\usepackage{jheppub} % for details on the use of the package, please see the JHEP-author-manual
\usepackage{tabularx} % extra features for tabular environment
\usepackage{amsmath,amsfonts}
\usepackage{dsfont}  % improve math presentation
\usepackage{graphicx} % takes care of graphic including machinery
\usepackage{slashed}
\usepackage{tikz-feynman}
\usepackage{mathtools}
\usepackage{import}
\usepackage[font=small,
            width=0.8\textwidth]{caption}
\hypersetup{
	colorlinks=true,       % false: boxed links; true: colored links
	linkcolor=black,        % color of internal links
	citecolor=blue,        % color of links to bibliography
	filecolor=magenta,     % color of file links
	urlcolor=black         
}
\usepackage{blindtext}

\DeclarePairedDelimiter{\floor}{\lfloor}{\rfloor}
\NewDocumentCommand{\evalat}{sO{\big}mm}{%
  \IfBooleanTF{#1}
   {\mleft. #3 \mright|_{#4}}
   {#3#2|_{#4}}%
}

\title{
    Effective ALP-Photon Coupling in External Magnetic Fields}
\author{O. Semin}
\affiliation{Institute for Theoretical Physics, Auf der Morgenstelle 14, University of Tübingen, Tübingen 72076, Germany}
\emailAdd{ozan.semin@uni-tuebingen.de}

\abstract{
We presented a complete calculation of the one-loop fermionic correction to the effective coupling between axion-like particles (ALPs) and photons within a constant, homogeneous magnetic field of arbitrary strength. This interaction, responsible for the Primakoff effect, is central to detecting axion-like particles in astrophysical settings and terrestrial experiments like helioscopes and haloscopes. Accurately predicting the interaction rate requires accounting for quantum corrections. Our work tackles this by employing magnetically field-dressed fermion propagators derived using Schwinger's proper time method and a systematic Lorentz decomposition using the Ritus basis. We evaluate the triangle loop diagram exactly, and compare it to approximations on field strength under specific assumptions.}

\begin{document}
\maketitle
\section{Introduction}
The Standard Model (SM) of particle physics, despite its remarkable success in describing fundamental particles and their interactions, falls short of describing the full picture. Among the most pressing is the nature of dark matter, which constitutes the vast majority of matter in the universe \cite{Zwicky1933Die,1980ApJ238471R,Massey_2010}. Its existence is inferred from overwhelming gravitational evidence, yet its composition remains uncertain, pointing towards physics Beyond the Standard Model (BSM) \cite{saikumar2024exploringfrontierschallengestheories,article,Gehrlein:2019iwl,PhysRevD.100.035041}. This gap in our understanding motivates a wide-ranging theoretical and experimental search for new particles.

Among the most promising BSM candidates are axions and axion-like particles (ALPs) \cite{Peccei2008}. The original QCD axion emerged as an elegant solution to the strong CP problem inherent in Quantum Chromodynamics (QCD), while ALPs arise more generally in various theoretical extensions of the SM, including string theory \cite{PhysRevD.81.123530}. These light, pseudoscalar bosons are not only theoretically well-motivated but also serve as viable dark matter candidates. A key phenomenological feature of ALPs is their potential coupling to photons, often parameterized by the coupling constant $g_{a\gamma\gamma}$. This interaction opens a window for their detection, as it mediates the conversion of ALPs into photons (and vice versa) in the presence of external electromagnetic fields, a phenomenon known as the Primakoff effect \cite{PhysRev.81.899}.

The study of ALP-photon conversion has been a very fruitful topic in different scales of physics. Astrophysical environments hosting extremely strong magnetic fields, such as magnetars and neutron stars, provide natural laboratories to probe this coupling \cite{Fortin_2018,Fortin_2021}. Terrestrial experiments employ strong laboratory magnets, including helioscopes like CAST and the future IAXO \cite{vogel2013iaxointernationalaxion,andriamonje2007improved}, which search for ALPs produced in the Sun, and haloscopes like ADMX and MADMAX \cite{PhysRevLett.118.091801,PhysRevLett.124.101303,PhysRevLett.120.151301}, designed to detect relic ALPs constituting the galactic dark matter halo.

Precise theoretical predictions are crucial for interpreting the results of these searches and guiding future experimental design. While the tree-level Primakoff effect provides a leading-order description, quantum corrections can significantly modify the effective ALP-photon coupling, especially in the strong field regimes relevant astrophysically and experimentally \cite{PhysRevLett.51.1415,PhysRevD.37.1237,PhysRevD.66.043517}. These corrections arise from loop diagrams involving charged particles that interact with both the ALP and the external magnetic field. Calculating these effects requires a non-perturbative treatment of the charged particle propagators within the background field. This is typically achieved within the Furry picture \cite{PhysRev.81.115}, where the interaction with the external field is incorporated exactly into the propagators using techniques like Schwinger's proper time method \cite{PhysRev.82.664}.

While the concept is established, obtaining exact analytical results for loop diagrams in arbitrary, constant magnetic fields remains a technical challenge, particularly for three-point functions like the ALP-photon-photon vertex. Previous studies often relied on approximations or specific kinematic limits. This work presents a complete calculation of the one-loop fermionic correction to the effective ALP-photon coupling in an arbitrarily strong, constant, and homogeneous magnetic field. We employ field-dressed fermion propagators mentioned above and perform the loop integration without resorting to approximations concerning the field strength or the kinematics of the external particles (allowing them to be on-shell or off-shell).

The paper is structured as follows: Section \ref{sec:theory} provides the necessary theoretical background on axions, ALPs, and their effective interactions. In Section~\ref{sec:lorentz}, we decompose the general Lorentz structure of the ALP-photon-photon vertex in the presence of an external magnetic field. Section \ref{sec:loop_calc} presents the core calculation of the one-loop triangle diagram, detailing the integration techniques used to handle the Schwinger phase and the proper-time integrals. Afterwards, in Section~\ref{sec:special_cases}, we compute both the fieldless and the strong field limit, latter assuming certain kinematic configurations. We conclude by summarizing our findings and their implications, and compare our results with the approximations. The explicit results of the trace calculations and further details are provided in a supplementary \textsc{GitHub} repository \footnote{\href{https://github.com/Ozzywtlk/Effective-ALP-Photon-Coupling-in-External-Magnetic-Fields}{https://github.com/Ozzywtlk/Effective-ALP-Photon-Coupling-in-External-Magnetic-Fields}}.

\section{Axions and ALPs}\label{sec:theory}
The theoretical motivation for axions stems from the strong CP problem in quantum chromodynamics (QCD). This problem arises from the potential inclusion of a CP-violating $\bar{\theta}$-term in the QCD Lagrangian \cite{d370fac2b94a4edca23e7cf28a022b27},  
\begin{equation}\label{eq:qcd_lagrangian}
    \mathcal{L}_{\text{QCD}}\supset \bar{\theta}\frac{\alpha_s}{8\pi}G^A_{\mu\nu}\tilde{G}_A^{\mu\nu},
\end{equation}  
where $\alpha_s$ is the strong coupling constant, $G_{\mu\nu}^A$ is the gluon field strength tensor, $\tilde{G}_A^{\mu\nu}$ is its dual, and $\bar{\theta}$ is an angular parameter. This term contributes to the neutron's electric dipole moment (EDM), but experimental bounds \cite{PhysRevLett.124.081803} constrain $|\bar{\theta}| < 10^{-10}$, implying an unexpectedly small value and posing a fine-tuning problem.

To render the $\bar{\theta}$ parameter physically redundant, Peccei and Quinn proposed introducing a new, anomalous global symmetry, denoted $U(1)_\text{PQ}$ \cite{1977PhRvL..38.1440P}. At high energies, this symmetry is postulated to act trivially on fields, leaving the Lagrangian invariant up to the topological term seen in Eq.~\eqref{eq:qcd_lagrangian}. However, consistency requires considering the low-energy implications. According to anomaly matching conditions, an anomalous global symmetry like $U(1)_\text{PQ}$ implies the existence of colored massless fermions in the low-energy theory \cite{tHooft:1980xss}. As such particles are experimentally ruled out, the $U(1)_\text{PQ}$ symmetry must be spontaneously broken at some high energy scale, $f_a$, typically via the vacuum expectation value of a complex scalar field. This spontaneous symmetry breaking results in a massless Goldstone boson $a$, the axion, which possesses a characteristic shift symmetry. Crucially, the original $U(1)_\text{PQ}$ symmetry suffers from an axial anomaly due to non-perturbative QCD effects \cite{PhysRev.177.2426}. This means the associated chiral current, $J^\mu_{\text{PQ}}$, is not conserved:
\begin{equation}
    \partial_\mu J^\mu_{\text{PQ}} = \frac{\alpha_s}{8\pi}G^A_{\mu\nu}\tilde{G}_A^{\mu\nu}.
\end{equation}
This anomaly directly generates an effective interaction between the axion and gluons:
\begin{equation}
    \mathcal{L}_{agg}= \frac{a}{f_a}\frac{\alpha_s}{8\pi}G^A_{\mu\nu}\tilde{G}_A^{\mu\nu}.
\end{equation}
When this term is added to the Lagrangian, the $\bar{\theta}$ parameter is effectively promoted to a dynamical field. Non-perturbative QCD effects generate a potential for the axion field, whose minimum dynamically cancels the original $\bar{\theta}$ term, thus solving the strong CP problem. This mechanism also generates a small mass for the axion, making it a \textit{pseudo-Goldstone boson}, related to the scale $f_a$ and quark masses \cite{Peccei2008}. Specifically, one obtains a mass term of the form\cite{Peccei2008} 
\begin{equation}
    m_a=-\frac{1}{f_a}\frac{\alpha_s}{4\pi}\frac{\partial}{\partial a}\langle G^A_{\mu\nu}\tilde{G}_A^{\mu\nu}\rangle\big|_{\langle a\rangle=-f_a\bar{\theta}}
\end{equation}
where the right hand side corresponds to the second derivative of the effective potential. The precise properties of the axion depend on the considered model, with the best-known constructions being the KSVZ\cite{PhysRevLett.43.103,SHIFMAN1980493} and DFSZ\cite{DINE1981199,osti_7063072} models. The KSVZ model introduces vector-like quarks that are charged under $U(1)_{\text{PQ}}$, while the DFSZ model introduces an additional Higgs doublet. A comprehensive review of the QCD axion can be found in \cite{di2016qcd}.

Astrophysical observations (e.g., neutron star cooling) and cosmological considerations constrain the allowed range for $f_a$, typically placing it between $\sim 10^9$ GeV and $\sim 10^{18}$ GeV \cite{PhysRevD.98.103015,Hannestad:2005df,KAWASAKI2018181}. This motivates the study of more general Axion-Like Particles (ALPs), where the mass $m_a$ and the effective scale $\Lambda$ (analogous to $f_a$) are treated as independent parameters. ALPs encompass the QCD axion but allow for a broader parameter space.

In an effective field theory (EFT) approach below the scale $\Lambda$, the leading interactions of ALPs involving SM fields (before electroweak symmetry breaking, EWSB) are often described by dimension-five operators \cite{georgi1986manifesting}:
\begin{equation}\label{eq:dim5operators}
    \mathcal{L}_\text{eff}^5=\frac{\partial^\mu a}{\Lambda}\sum_f \bar{\psi}_fC_{ff}\gamma^5\gamma_\mu\psi_f+\frac{a}{\Lambda}\left(g_s^2C_{GG}G_{\mu\nu}^A\tilde{G}^{\mu\nu,A}+g^2C_{WW}W_{\mu\nu}^A\tilde{W}^{\mu\nu,A} +g^{\prime 2}C_{BB}B_{\mu\nu}\tilde{B}^{\mu\nu}\right).
\end{equation}
Here, $\psi_f$ represents SM fermion fields, $G^{\mu\nu}$, $W^{\mu\nu}$, and $B^{\mu\nu}$ are the field strength tensors for $SU(3)_C$, $SU(2)_L$, and $U(1)_Y$ respectively, and $C_{ff}, C_{GG}, C_{WW}, C_{BB}$ are model-dependent Wilson coefficients. The derivative coupling to fermions respects the approximate shift symmetry $a \to a + c$ expected for a pseudo-Goldstone boson. Below the EWSB scale, the $W$ and $B$ interactions mix, generating an effective coupling to photons:
\begin{equation}\label{eq:gammagamma}
    \mathcal{L}_{a\gamma\gamma} = \frac{g_{a\gamma\gamma}}{4} a F_{\mu\nu} \tilde{F}^{\mu\nu}, \quad \text{with} \quad g_{a\gamma\gamma}=4\frac{e^2}{\Lambda}(C_{BB}+C_{WW}) \equiv 4\frac{e^2}{\Lambda}C_{\gamma\gamma}.
\end{equation}
This ALP-photon coupling $g_{a\gamma\gamma}$, typically treated as a free parameter and is constrained by experimental data, is the primary focus of experimental searches based on the Primakoff effect and is the central quantity investigated in this paper, including its modification by quantum effects in external magnetic fields. By performing loop calculations, one can derive the effective coupling constants $g_{aX}^\text{eff.}$, whereas the exact definition of \textit{effective} depends on the context of the calculation. For instance, in the context of ALP to photon differential decay rate, the effective coupling can be defined using
\begin{equation}
    \frac{\text{d}\Gamma}{\text{d}\Omega}=\frac{m_a^3|g_{a\gamma\gamma}^\text{eff.}|^2}{128\pi^2},
\end{equation}
such that $|g_{a\gamma\gamma}^\text{eff.}|$ is extracted after the loop calculation. In the following sections, we will be using this to compare our results, yet our calculations are performed in a completely general way, such that the effective coupling can be defined in any context.

\section{Lorentz Structure of the ALP-Photon-Photon Vertex}\label{sec:lorentz}
\begin{figure}[!htb]
    \centering    
    \includegraphics[width=0.6\textwidth]{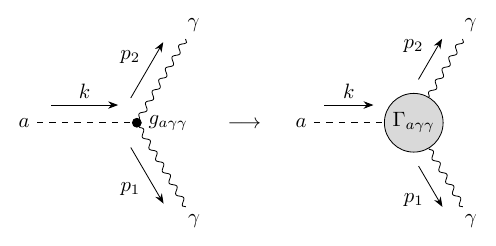}
    \caption{ALP-photon-photon coupling with (right) and without (left) external magnetic field effects.}
    \label{fig:blob}
\end{figure}

Starting from the tree-level ALP-photon interaction Lagrangian in Eq.~\eqref{eq:gammagamma}, characterized by the coupling $g_{a\gamma\gamma}$, we calculate the modifications induced by an external magnetic field. These modifications arise from one-loop diagrams involving fermions interacting non-perturbatively with the background field. This leads to an effective vertex function, $\Gamma_{a\gamma\gamma}^{\mu\nu\rho}$, which manifests the field-dependent corrections, replacing the simple point-like interaction, as illustrated in Figure~\ref{fig:blob}.
where the ALP momentum is denoted as $k$, while momenta of the two photons are $p_i$. 
To analyze the field-dressed coupling $\Gamma_{a\gamma\gamma}^{\mu\nu\rho}$, we begin by decomposing it into a basis of Lorentz structures. An important analytical property of this tensor, similar to the QED polarization tensor in an external ﬁeld, is that it is non‐Hermitian. This arises because the intermediate fermions in the loop can go on‐shell, corresponding to physical processes like pair production. The resulting imaginary part of the vertex function describes the rate of such absorptive phenomena.

This decomposition simplifies the calculation, as the coefficients associated with each structure can be isolated by contracting $\Gamma_{a\gamma\gamma}^{\mu\nu\rho}$ with the corresponding basis element, provided the basis is chosen to be orthogonal. The fundamental building blocks for these structures are the external momenta $k$, $p_1$, and $p_2$. While various orthogonalization schemes exist, the presence of an external magnetic field makes it advantageous to incorporate the field strength tensor $F_\text{c}^{\mu\nu}$ into the basis construction. The general Lorentz structure of the vertex can then be built from linear combinations of the following four-vectors:
\begin{equation}
    q^\mu,~~F_\text{c}^{\mu\nu}q_\nu,~~ F_\text{c}^{\mu\nu}F_{c,\nu\rho} q^\rho,~~ \tilde{F}_\text{c}^{\mu\nu}q_\nu,
\end{equation}
where $q\in\{k,p_1,p_2\}$. Additional contractions with the dual field strength tensor are omitted, as they either reduce to the above structures or vanish due to the absence of external electric fields.\footnote{In other Lorentz frames, acquired electric fields would be orthogonal to the magnetic field, satisfying the Lorentz-invariant cross-field condition, $\vec{E}\dot\vec{B}=0$.}

To ensure mutual orthogonality and completeness, we work in the Ritus basis for the Lorentz structures \cite{papanjan1971vacuum,papanyan1973three}, which consists of the vectors
\begin{equation}
    q^\mu,~~ L_q^\mu := \hat{F}_\text{c}^{\mu\nu} q_\nu,~~ \tilde{L}_q^\mu := \hat{\tilde{F}}_\text{c}^{\mu\nu} q_\nu,~~ G_q^\mu := \frac{q^2 \hat{F}_\text{c}^{\mu\nu} \hat{F}_{c,\nu\rho} q^\rho}{L_q^2} + q^\mu.
\end{equation}
For on-shell photons ($p_i^2=0$), the vector $G_{p_i}^\mu$ simplifies to $p_i^\mu$. Geometrically, these basis vectors can be drawn as projections of the external momenta parallel and perpendicular to the magnetic field, analogous to polarization vectors. For the pseudoscalar ALP, however, this basis decomposition serves primarily as a mathematical tool.

We assign the first index ($\mu$) of the vertex tensor $\Gamma_{a\gamma\gamma}^{\mu\nu\rho}$ to the ALP's axial current and the remaining indices ($\nu, \rho$) to the photons. allowing us to impose the Ward identities for the photons:
\begin{equation}
    \Gamma_{a\gamma\gamma}^{\mu\nu\rho}(k,p_1,p_2) p_{1,\nu} = 0,\quad \Gamma_{a\gamma\gamma}^{\mu\nu\rho}(k,p_1,p_2) p_{2,\rho} = 0.
\end{equation}
These identities ensure that the decomposition contains no terms proportional to $p_1^\nu$ or $p_2^\rho$. Additionally, Bose symmetry requires the vertex to be symmetric under the exchange of the two photons ($p_{1, \nu} \leftrightarrow p_{2, \rho}$). Incorporating these constraints, the general expansion of this third-rank pseudotensor with Bose symmetry in the Ritus basis is:
\begin{align}
    \Gamma_{a\gamma\gamma}^{\mu\nu\rho}&=
    c_1\hat{k}^\mu \hat{L}_{p_1}^\nu \hat{L}_{p_2}^\rho +
    c_2\hat{k}^\mu \hat{\tilde{L}}_{p_1}^\nu \hat{\tilde{L}}_{p_2}^\rho +
    c_3\hat{k}^\mu \hat{G}_{p_1}^\nu \hat{G}_{p_2}^\rho +
    c_4\hat{L}_k^\mu \hat{L}_{p_1}^\nu \hat{L}_{p_2}^\rho +
    c_5\hat{L}_k^\mu \hat{\tilde{L}}_{p_1}^\nu \hat{\tilde{L}}_{p_2}^\rho +
    c_6\hat{L}_k^\mu \hat{G}_{p_1}^\nu \hat{G}_{p_2}^\rho 
    \nonumber\\&+
    c_7\hat{\tilde{L}}_k^\mu \hat{L}_{p_1}^\nu \hat{L}_{p_2}^\rho +
    c_8\hat{\tilde{L}}_k^\mu \hat{\tilde{L}}_{p_1}^\nu \hat{\tilde{L}}_{p_2}^\rho +
    c_9\hat{\tilde{L}}_k^\mu \hat{G}_{p_1}^\nu \hat{G}_{p_2}^\rho +
    c_{10}\hat{G}_k^\mu \hat{L}_{p_1}^\nu \hat{L}_{p_2}^\rho +
    c_{11}\hat{G}_k^\mu \hat{\tilde{L}}_{p_1}^\nu \hat{\tilde{L}}_{p_2}^\rho 
    \nonumber\\&+
    c_{12}\hat{G}_k^\mu \hat{G}_{p_1}^\nu \hat{G}_{p_2}^\rho +
    \frac{c_{13}}{\sqrt{2}}\hat{k}^\mu\left(\hat{L}_{p_1}^\nu \hat{\tilde{L}}_{p_2}^\rho+\hat{\tilde{L}}_{p_1}^\nu \hat{L}_{p_2}^\rho\right)+
    \frac{c_{14}}{\sqrt{2}}\hat{k}^\mu\left(\hat{L}_{p_1}^\nu \hat{G}_{p_2}^\rho+\hat{G}_{p_1}^\nu \hat{L}_{p_2}^\rho\right)
    \nonumber\\&+
    \frac{c_{15}}{\sqrt{2}}\hat{k}^\mu\left(\hat{\tilde{L}}_{p_1}^\nu \hat{G}_{p_2}^\rho+\hat{G}_{p_1}^\nu \hat{\tilde{L}}_{p_2}^\rho\right)+
    \frac{c_{16}}{\sqrt{2}}\hat{L}_k^\mu\left(\hat{L}_{p_1}^\nu \hat{\tilde{L}}_{p_2}^\rho+\hat{\tilde{L}}_{p_1}^\nu \hat{L}_{p_2}^\rho\right)+
    \frac{c_{17}}{\sqrt{2}}\hat{L}_k^\mu\left(\hat{L}_{p_1}^\nu \hat{G}_{p_2}^\rho+\hat{G}_{p_1}^\nu \hat{L}_{p_2}^\rho\right)
    \nonumber\\&+
    \frac{c_{18}}{\sqrt{2}}\hat{L}_k^\mu\left(\hat{\tilde{L}}_{p_1}^\nu \hat{G}_{p_2}^\rho+\hat{G}_{p_1}^\nu \hat{\tilde{L}}_{p_2}^\rho\right)+
    \frac{c_{19}}{\sqrt{2}}\hat{\tilde{L}}_k^\mu\left(\hat{L}_{p_1}^\nu \hat{\tilde{L}}_{p_2}^\rho+\hat{\tilde{L}}_{p_1}^\nu \hat{L}_{p_2}^\rho\right)+
    \frac{c_{20}}{\sqrt{2}}\hat{\tilde{L}}_k^\mu\left(\hat{L}_{p_1}^\nu \hat{G}_{p_2}^\rho+\hat{G}_{p_1}^\nu \hat{L}_{p_2}^\rho\right)
    \nonumber\\&+
    \frac{c_{21}}{\sqrt{2}}\hat{\tilde{L}}_k^\mu\left(\hat{\tilde{L}}_{p_1}^\nu \hat{G}_{p_2}^\rho+\hat{G}_{p_1}^\nu \hat{\tilde{L}}_{p_2}^\rho\right)+
    \frac{c_{22}}{\sqrt{2}}\hat{G}_k^\mu\left(\hat{L}_{p_1}^\nu \hat{\tilde{L}}_{p_2}^\rho+\hat{\tilde{L}}_{p_1}^\nu \hat{L}_{p_2}^\rho\right)+
    \frac{c_{23}}{\sqrt{2}}\hat{G}_k^\mu\left(\hat{L}_{p_1}^\nu \hat{G}_{p_2}^\rho+\hat{G}_{p_1}^\nu \hat{L}_{p_2}^\rho\right)
    \nonumber\\&+
    \frac{c_{24}}{\sqrt{2}}\hat{G}_k^\mu\left(\hat{\tilde{L}}_{p_1}^\nu \hat{G}_{p_2}^\rho+\hat{G}_{p_1}^\nu \hat{\tilde{L}}_{p_2}^\rho\right)
\end{align}\label{eq:vertex_decomposition}
Here, hats denote normalized vectors, e.g., $\hat{q}^\mu = q^\mu / \sqrt{|q^2|}$. Note that the calculation of these coefficients is not always necessary, but useful, especially when the result itself is lengthy. Without specific assumptions about the CP properties governing the ALP interactions, this general decomposition cannot be reduced further. We now proceed to the explicit calculation of the one-loop diagram that determines the coefficients $c_i$.

\section{Calculation of The Field-Dressed Triangle Loop}\label{sec:loop_calc} 
\begin{figure}[!htb]
    \centering
    \includegraphics[width=0.7\textwidth]{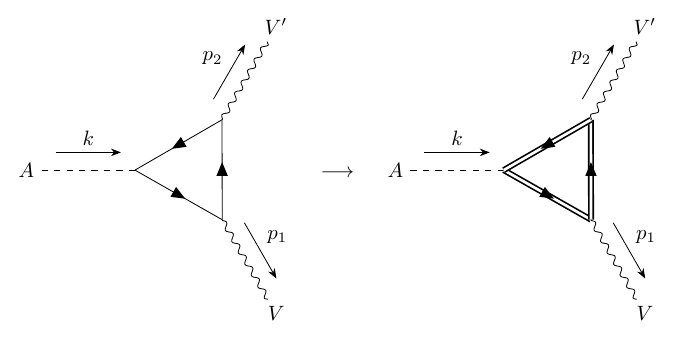}
    \caption{Triangle loop diagram for the vector-vector-axial vector coupling under the influence of external magnetic fields.}
    \label{fig:axion_photon_conversion}
\end{figure}

In our calculation, we treat both photons in the ALP–photon–photon vertex as quantum fields, rather than considering one as a classical external field (as is often done in similar calculations). This means we do not use a semiclassical approximation for the external magnetic field. Since there are no tree-level diagrams for ALP–photon conversion, in order to account for how the external magnetic field modifies this interaction, we must consider one-loop diagrams mediated by electromagnetically charged particles. Focusing on ALP–fermion interactions, the corresponding Feynman rule for the ALP-fermion interaction vertex in the momentum space can be derived from Eq.~(\ref{eq:dim5operators}) as
\begin{equation}\label{eq:vertex}
    \vcenter{\hbox{\includegraphics[width=0.35\textwidth]{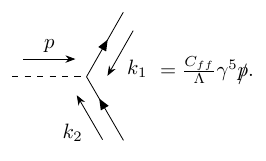}}}
\end{equation}
The one-loop contribution to the vertex is given by the sum of two triangle diagrams. The first is depicted in Figure~\ref{fig:axion_photon_conversion}, and the second is identical but with the two external photon legs exchanged ($p_{1, \nu} \leftrightarrow p_{2, \rho}$). As mentioned above, Bose symmetry requires that the total amplitude be symmetric under this exchange. For brevity, only one diagram is shown, but our results at the end are explicitly constructed as a Bose-symmetric vertex $\Gamma_{a\gamma\gamma}^{\mu\nu\rho}$ with both of the contributions.
\\
For the integration over the internal loop momenta, we employ the methodology presented in \cite{PhysRevD.107.116024,Ayala_2024}, designed to obtain exact results for arbitrary magnetic field strengths. The core strategy is to evaluate the spacetime integrals first, followed by the momentum integrals, which are brought into a Gaussian form. Our calculation focuses on the vector-vector-pseudovector loop diagram, involving the shift-invariant fermion-pseudovector vertex given in Eq.~\eqref{eq:vertex}. We define the constant magnetic field direction to be parallel to the $z$-axis.

We begin by examining the properties of the phase factor $\Phi(x,x')$ from Eq.~\eqref{eq:phase_factor}. The integral defining $\Phi(x,x')$ is path-independent. Consequently, for a path traversing from $x'$ to $x$ and then back to $x'$, the total phase accumulation is zero:
\begin{equation}
    \Phi(x,x^\prime)+\Phi(x^\prime,x)=-\frac{e_f}{2}F^\text{c}_{\mu\nu}(x-x^\prime)^\nu\int\limits_{x^\prime}^x\text{d}\xi^\mu=-\frac{e_f}{2}F_{c,\mu\nu}(x-x^\prime)^\nu(x-x^\prime)^\mu=0.
\end{equation}
For a closed loop integral involving $n \geq 3$ vertices, the total Schwinger phase factor $\Phi_\Sigma$ generally does not vanish. As discussed in the Appendix~\ref{app:propagators}, each a phase factor $\Phi(x, x')$ depends on the gauge choice and the endpoints of the internal fermion line. Therefore when combining propagators in a loop, these individual phases do not necessarily cancel. In the specific case of the triangle loop ($n=3$) relevant here, with vertices at $x, y, z$, the total phase factor is the sum of the phases along each leg: $\Phi_\Sigma(x,y,z) = \Phi(x,y) + \Phi(y,z) + \Phi(z,x)$. To evaluate this phase explicitly, we use the gauge freedom to choose a convenient representation for the constant magnetic field's vector potential $A^\text{c}$:
\begin{equation}
    A^\text{c}=\frac{|B|}{2}(0,-y,x,0).
\end{equation}
In this gauge, the total phase accumulated around the loop defined by vertices at $x, y, z$ is
\begin{equation}
    \Phi_\Sigma(x,y,z):=-\frac{\beta }{2}\left(x_\mu \hat{F}_\text{c}^{\mu\nu}y_\nu+y_\mu \hat{F}_\text{c}^{\mu\nu}z_\nu+z_\mu \hat{F}_\text{c}^{\mu\nu}x_\nu\right).\label{eq:phase}
\end{equation}
Since this expression is generally non-zero, the manifest translation invariance (and thus momentum conservation) is obscured in this coordinate-space representation. It proves more convenient to work with the translation-invariant part of the propagator, Eq.~(\ref{eq:fermion_propagator}), expressed in momentum space:
\begin{align}
    S_F(p)&=\int\limits_0^\infty\frac{\text{d}s_j}{\mathfrak{c}_j}\left[\left(m_f+\slashed{p}_\parallel\right)\mathfrak{e}_j+\frac{\slashed{p}_\perp}{\mathfrak{c}_j}\right]e^{-is\left(m_f^2-p_\parallel^2-p_\perp^2\frac{\mathfrak{t}_j}{\beta s}\right)},\label{eq:fermion_propagator_fourier}
\end{align}
where $p_\parallel:=(p^0,0,0,p^3)$ and $p_\perp:=(0,p^1,p^2,0)$ denote the components parallel and perpendicular to the magnetic field, respectively. Note that this decomposition differs from the Ritus basis vectors $L_p$ and $\tilde{L}_p$ due to the index structure arising from the field strength tensor's antisymmetry. For brevity, we introduce the following abbreviations for trigonometric and exponential factors dependent on the proper-time variable $s_j$:
\begin{align}
    \mathfrak{t}_j&:=\tan(\beta s_j)\nonumber\\
    \mathfrak{c}_j&:=\cos(\beta s_j)\nonumber\\
    \mathfrak{s}_j&:=\sin(\beta s_j)\nonumber\\
    \mathfrak{e}_j&:=e^{i\beta s_j\Sigma_3}=\mathfrak{c_j}+i\mathfrak{s_j}\Sigma_3.
\end{align} 
where $\Sigma_3$ corresponds to the spin operator along the magnetic field and $\beta=e_f B$.

With these considerations, we express the tensor-valued vertex loop integral as
\begin{align}
    \Gamma^{\mu\nu\rho}_{a\gamma\gamma}(k,p_1,p_2)&=-\frac{i}{2}\sum\limits_f\frac{e_f^2C_{ff}}{\Lambda}
    \iiint\frac{\text{d}^4q}{(2\pi)^4}~\frac{\text{d}^4\ell}{(2\pi)^4}~\frac{\text{d}^4w}{(2\pi)^4}
    \iiint\text{d}^4x~\text{d}^4y~\text{d}^4z\nonumber\\
    &\times e^{i\Phi_\Sigma(x,y,z)}\text{Tr}\left[\gamma^5\gamma^\mu S_f(q)\gamma^\nu S_f(\ell)\gamma^\rho S_f(w)\right]\nonumber\\
    &\times e^{-ix(k+q-w)-iy(-p_1+w-\ell)-iz(-p_2+\ell-q)}.
\end{align}
The spacetime integrals originate from the interaction vertices, while the momentum integrals arise from the Fourier transformation of the fermion propagators.
\subsection{Integration over the Spacetime Coordinates}
We note that the quantities in the phase term are all contracted with the field strength tensor, meaning that only components perpendicular to the magnetic field remain in the exponential. Therefore, we decompose the spacetime measure as 
\begin{equation}
    \text{d}^4x=\text{d}^{2}x_\perp\text{d}^{2}x_\parallel,
\end{equation}   
and perform the integration over $x_\parallel$, $y_\parallel$, and $z_\parallel$. This yields
\begin{align}
    \Gamma^{\mu\nu\rho}_{a\gamma\gamma}(k,p_1,p_2)&=-\frac{i(2\pi)^{6}}{2}\sum\limits_f\frac{e_f^2C_{ff}}{\Lambda}
    \iiint\frac{\text{d}^4q}{(2\pi)^4}~\frac{\text{d}^4\ell}{(2\pi)^4}~\frac{\text{d}^4w}{(2\pi)^4}
    \delta^{\left(2\right)}(k_\parallel+q_\parallel-w_\parallel)\nonumber\\
    &\times \delta^{\left(2\right)}(-p_{1\parallel}+w_\parallel-\ell_\parallel)\delta^{\left(2\right)}(-p_{2\parallel}+\ell_\parallel-q_\parallel)\text{Tr}\left[\gamma^5\gamma^\mu S_f(q)\gamma^\nu S_f(\ell)\gamma^\rho S_f(w)\right]\nonumber\\
    &\times\underbrace{\iiint\text{d}^{2}x_\perp~\text{d}^{2}y_\perp~\text{d}^{2}z_\perp e^{i[\Phi_\Sigma(x,y,z)+K_1(x,y,z)]} }_{=:A},\label{eq:vertex_loop}
\end{align}
with
\begin{equation}
    K_1(x,y,z)=-x_\perp(k_\perp+q_\perp-w_\perp)-y_\perp(-p_{1\perp}+w_\perp-\ell_\perp)-z_\perp(-p_{2\perp}+\ell_\perp-q_\perp)\label{eq:other_phase}
\end{equation}
where momentum conservation along the magnetic field is enforced by the delta functions. The remaining part of the integration requires additional evaluation, which we perform separately for each perpendicular variable. In this specific magnetic field configuration, the action of $\hat{F}_\text{c}$ on spacetime coordinates projects them onto the $xy$-plane while simultaneously rotating them by $90^\circ$ around the $z$-axis, since $\hat{F}_{c}^{12}=-\hat{F}_{c}^{21}=1$, with all other matrix components vanishing. This means that on the subspace perpendicular to the magnetic field, we have
\begin{equation}
    \evalat[\Big]{\hat{F}_{c}^2}{\perp}=-\mathds{1}_{\perp}.
\end{equation}
Using this property, we rewrite the phase terms in Eqs.~\eqref{eq:phase} and \eqref{eq:other_phase} as 
\begin{align}
    \Phi_\Sigma(x,y,z)&=-\frac{\beta }{2}\left(x_\perp\hat{F}_\text{c}\left[y_\perp-z_\perp\right]+y_\perp \hat{F}_\text{c} z_\perp\right)
    \\[1mm]
    K_1(x,y,z)&=x_\perp\hat{F}_\text{c}^2(k_\perp+q_\perp-w_\perp)\underbrace{-y_\perp(-p_{1\perp}+w_\perp-\ell_\perp)-z_\perp(-p_{2\perp}+\ell_\perp-q_\perp)}_{=:K_2(y,z)}.
\end{align}
After performing the rotation $x_\perp \hat{F}_\text{c}\to\tilde{x}_\perp$, which leaves the measure invariant, we can perform the integration over the $\tilde{x}_\perp$ variable and obtain 
\begin{align}
    A&=\left(\frac{4\pi}{\beta }\right)^2\iint\text{d}^{2}y_\perp~\text{d}^{2}z_\perp\delta^{(2)}\left(y_\perp-z_\perp-\frac{2\hat{F}_\text{c}}{\beta }[k_\perp+q_\perp-w_\perp]\right)e^{-i\frac{\beta }{2}y_\perp\hat{F}_\text{c} z_\perp+iK_2(y,z)}\nonumber\\
    &=\left(\frac{4\pi}{\beta }\right)^2\int\text{d}^{2}z_\perp e^{-i[k_\perp-p_{1,\perp}-p_{2,\perp}]z_\perp+i[k_\perp+q_\perp-w_\perp]\frac{2\hat{F}_\text{c}}{\beta }[-p_{1\perp}+w_\perp-\ell_\perp]}\nonumber\\
    &=\left(\frac{8\pi^2}{\beta }\right)^2\delta^{(2)}\left(k_\perp-p_{1,\perp}-p_{2,\perp}\right)e^{i[k_\perp+q_\perp-w_\perp]\frac{2\hat{F}_\text{c}}{\beta }[-p_{1\perp}+w_\perp-\ell_\perp]},
\end{align}
where we have used $z_\perp \hat{F}_\text{c} z_\perp=0$ in the second step. Inserting this result into Eq.~\eqref{eq:vertex_loop}, and performing integration over the parallel momenta, we obtain
\begin{align}
    \Gamma^{\mu\nu\rho}_{a\gamma\gamma}(k,p_1,p_2)&=-2\frac{i}{\left(2\pi  \right)^2}\sum\limits_f\frac{e_f^2C_{ff}}{\beta^2\Lambda}
    \iiint\text{d}^4q~\text{d}^4\ell~\text{d}^4w~\delta^{\left(2\right)}(k_\parallel+q_\parallel-w_\parallel)\nonumber\\
    &\times \delta^{\left(2\right)}(-p_{1\parallel}+w_\parallel-\ell_\parallel)\delta^{\left(2\right)}(-p_{2\parallel}+\ell_\parallel-q_\parallel)\text{Tr}\left[\gamma^5\gamma^\mu S_f(q)\gamma^\nu S_f(\ell)\gamma^\rho S_f(w)\right]\nonumber\\ 
    &\times\delta^{(2)}\left(k_\perp-p_{1,\perp}-p_{2,\perp}\right)e^{i[k_\perp+q_\perp-w_\perp]\frac{2\hat{F}_\text{c}}{\beta }[-p_{1\perp}+w_\perp-\ell_\perp]}\nonumber\\
    &=-2\frac{i}{\left(2\pi \right)^2}\sum\limits_f\frac{e_f^2C_{ff}}{beta^2\Lambda}
    \iiint\text{d}^2q_\perp~\text{d}^2\ell_\perp~\text{d}^4w~\delta^{\left(4\right)}(k-p_{2}-p_{1})\nonumber\\
    &\times\text{Tr}\left[\gamma^5\gamma^\mu S_f\left(q\rvert_{q_\parallel=w_\parallel-k_{\parallel}}\right)\gamma^\nu S_f\left(\ell\rvert_{\ell_\parallel=w_\parallel-p_{1\parallel}}\right)\gamma^\rho S_f(w)\right]\nonumber\\
    &\times e^{i[k_\perp+q_\perp-w_\perp]\frac{2\hat{F}_\text{c}}{\beta }[-p_{1\perp}+w_\perp-\ell_\perp]}.\label{eq:vertex_loop_final}
\end{align}
Combined with the conservation of momentum along the magnetic field, this establishes the conservation of total momentum. This result is expected, as a uniformly constant magnetic field can be interpreted as a superposition of photons in different occupation number states with an infinite wavelength, i.e., zero momentum. 

\subsection{Integration over the Internal Momenta}
The last term in Eq.~\eqref{eq:vertex_loop_final} is the remnant of the Schwinger phase term. For clarity, we imply the conditions \(q_\parallel=w_\parallel-k_{\parallel}\) and \(\ell_\parallel=w_\parallel-p_{1\parallel}\) without writing them explicitly, unless stated otherwise. We can bring the integral above into a more convenient form by expanding the propagators and rewriting the order of the integrals. We then have
\begin{align}
    \Gamma^{\mu\nu\rho}_{a\gamma\gamma}(k,p_1,p_2)&=-2\frac{i}{\left(2\pi  \right)^2}\sum\limits_f\frac{e_f^2C_{ff}}{\beta^2\Lambda}\iiint\frac{\text{d}s_1~\text{d}s_2~\text{d}s_3}{\mathfrak{c_1}\mathfrak{c_2}\mathfrak{c_3}}\delta^{\left(4\right)}(k-p_{2}-p_{1})\nonumber\\
    &\times e^{-im_f^2(s_1+s_2+s_3)}\iiint\text{d}^2q_\perp~\text{d}^2\ell_\perp~\text{d}^4w~I^{\mu\nu\rho},\label{eq:vertex_loop_final_expanded}
\end{align}
with
\begin{align}
    I^{\mu\nu\rho}&=e^{i[k_\perp+q_\perp-w_\perp]\frac{2\hat{F}_\text{c}}{\beta }[-p_{1\perp}+w_\perp-\ell_\perp]} e^{is_1(w_\parallel-k_\parallel)^2+is_2(w_\parallel-p_{1,\parallel})^2+is_3w_\parallel^2}e^{\frac{i}{\beta }\left(q_\perp^2\mathfrak{t}_1+\ell_\perp^2\mathfrak{t}_2+w_\perp^2\mathfrak{t}_3\right)}\nonumber\\
    &\times \text{Tr}\left[\gamma^5\gamma^\mu \Pi_{1}(q)\gamma^\nu \Pi_{2}(\ell)\gamma^\rho \Pi_{3}(w)\right],
\end{align}
and
\begin{equation}
    \Pi_{j}(p):=(m_f+\slashed{p}_\parallel)\mathfrak{e}_j+\frac{\slashed{p}_\perp}{\mathfrak{c}_j}.
\end{equation}
We first perform the integration over the internal momenta instead of the proper time variable, and thus focus our attention on $I^{\mu\nu\rho}$. Starting with the $q$ variable, we obtain
\begin{align}
    \int\text{d}^2q_\perp~I^{\mu\nu\rho}&=C_{q_\perp}\int\text{d}^2q_\perp~e^{\frac{i}{\beta }\left(2q_\perp \hat{F}_\text{c}[w_\perp-p_{1\perp}-\ell_\perp]+q_\perp^2\mathfrak{t}_1\right)}\text{Tr}\left[\gamma^5\gamma^\mu \Pi_{1}(q)\gamma^\nu \Pi_{2}(\ell)\gamma^\rho \Pi_{3}(w)\right]\nonumber\\
    &=C_{q_\perp}e^{-i\frac{[w_\perp-p_{1\perp}-\ell_\perp]^2}{\beta \mathfrak{t_1}}}\int\text{d}^2q_\perp~e^{i\frac{q_\perp^2\mathfrak{t}_1}{\beta }}\nonumber\\
    &\times\text{Tr}\left[\gamma^5\gamma^\mu \Pi_{1}\left(q+\frac{\hat{F}_\text{c}}{\mathfrak{t}_1}\left[w_\perp-p_{1\perp}-\ell_\perp\right]\right)\gamma^\nu \Pi_{2}(\ell)\gamma^\rho \Pi_{3}(w)\right]\nonumber\\
    &=-\frac{i\pi \beta }{\mathfrak{t}_1}C_{q_\perp}e^{-i\frac{[w_\perp-p_{1\perp}-\ell_\perp]^2}{\beta \mathfrak{t_1}}}\nonumber\\
    &\times\text{Tr}\left[\gamma^5\gamma^\mu \Pi_{1}\left(q_\parallel+\frac{\hat{F}_\text{c}}{\mathfrak{t}_1}\left[w_\perp-p_{1\perp}-\ell_\perp\right]\right)\gamma^\nu \Pi_{2}(\ell)\gamma^\rho \Pi_{3}(w)\right],\label{eq:q_integral}
\end{align}
where we have shifted $q\to q+\frac{\hat{F}_\text{c}}{\mathfrak{t}_1}[w_\perp-p_{1\perp}-\ell_\perp]$ to bring the integral into a Gaussian form. The factor $C_{q_\perp}$ contains terms that are not affected by the $q_\perp$ integration.\par
An almost identical strategy is followed for $\ell_\perp$, yielding
\begin{align}
    \iint\text{d}^2\ell_\perp~\text{d}^2q_\perp I^{\mu\nu\rho}&=-\pi^2 \beta^2\frac{1}{\mathfrak{t}_1\mathfrak{t}_2-1}C_{\ell_\perp}e^{-\frac{i}{\beta }\frac{\mathfrak{t}_1}{\mathfrak{t}_1\mathfrak{t}_2-1}\left([w_\perp-k_\perp]\hat{F}_\text{c}+\frac{1}{\mathfrak{t}_1}[w_\perp-p_{1\perp}]\right)^2}\nonumber\\
    &\times\Bigg\{\text{Tr}\left[\begin{array}{l}
        \gamma^5\gamma^\mu \Pi_{1}\left(q_\parallel+\frac{\mathfrak{t}_2}{\mathfrak{t}_1\mathfrak{t}_2-1}\hat{F}_\text{c}[w_\perp-p_{1\perp}]-\frac{w_\perp-k_\perp}{\mathfrak{t}_1\mathfrak{t}_2-1}\right)\gamma^\nu\\
        \times\Pi_{2}\left(\ell_\parallel-\frac{1}{\mathfrak{t}_1\mathfrak{t}_2-1}\left[\mathfrak{t}_1\hat{F}_\text{c}[w_\perp-k_\perp]+[w_\perp-p_{1\perp}]\right]\right)\gamma^\rho \Pi_{3}(w) 
    \end{array}\right]\nonumber\\
    &-i\beta \frac{1}{2\mathfrak{c}_1\mathfrak{c}_2(\mathfrak{t}_1\mathfrak{t}_2-1)}\text{Tr}\left[\gamma^5\gamma^\mu\gamma_{\alpha}^\perp\hat{F}_\text{c}^{\alpha\beta}\gamma^\nu\gamma_{\perp\beta}\gamma^\rho\Pi_3(w)\right]\Bigg\}.
\end{align}
Here, the variable shift is $\ell\to\ell-\frac{\mathfrak{t}_1}{\mathfrak{t}_1\mathfrak{t}_2-1}(\hat{F}_\text{c}[w_\perp-k_\perp]+\frac{1}{\mathfrak{t}_1}[w_\perp-p_{1\perp}])$, and $C_{\ell_\perp}$ collects the terms unaffected by the $\ell_\perp$ integration. Due to the additional $\ell_\perp$ dependence in Eq.~\eqref{eq:q_integral}, we obtain a polynomial of order two in the $\ell_\perp$ variable, producing the latter trace in the result above.\par 
Since the integral over the $w$ variable is decoupled, we again separate the $w_\perp$ and $w_\parallel$ integrals. Although the $w_\perp$ integration follows a similar structure to the previous integrals, it is quite lengthy. Therefore, we present the final result for the integration over the entire $w$ variable:
\begin{align}
    \iiint\text{d}^4w~\text{d}^2\ell_\perp~\text{d}^2q_\perp I^{\mu\nu\rho}&=
    \frac{i\pi^4 \beta^3}{2}\frac{e^{\frac{i}{\beta }\frac{p_{1\perp}^2\mathfrak{t}_2\mathfrak{t}_3+p_{2\perp}^2\mathfrak{t}_1\mathfrak{t}_2+k_{\perp}^2\mathfrak{t}_1\mathfrak{t}_3+2\mathfrak{t}_1\mathfrak{t}_2\mathfrak{t}_3k_\perp\hat{F}_\text{c}p_{1\perp}}{\mathfrak{t}_1+\mathfrak{t}_2+\mathfrak{t}_3-\mathfrak{t}_1\mathfrak{t}_2\mathfrak{t}_3}}}{\mathfrak{t}_1+\mathfrak{t}_2+\mathfrak{t}_3-\mathfrak{t}_1\mathfrak{t}_2\mathfrak{t}_3}\nonumber
    \\
    &\times \frac{1}{s_1+s_2+s_3}e^{\frac{i}{s_1+s_2+s_3}\left(s_1s_3k_\parallel^2+s_2s_3p_{1\parallel}^2+s_1s_2p_{2\parallel}^2\right)}\nonumber\\
    &\times\left(\frac{i\beta \text{Tr}\left[\gamma^5\gamma^\mu\mathfrak{T}_1^{\nu\rho}\right]}{\mathfrak{t}_1+\mathfrak{t}_2+\mathfrak{t}_3-\mathfrak{t}_1\mathfrak{t}_2\mathfrak{t}_3}-\frac{i\text{Tr}\left[\gamma^5\gamma^\mu\mathfrak{T}_2^{\nu\rho}\right]}{s_1+s_2+s_3}+2\text{Tr}\left[\gamma^5\gamma^\mu\mathfrak{T}_3^{\nu\rho}\right]\right)\label{eq:operators}    
\end{align}
with the operators
\begin{align}
    \mathfrak{T}_1^{\nu\rho}&:=\frac{\eta_\perp^{\alpha\beta}+\mathfrak{t}_1\hat{F}_\text{c}^{\alpha\beta}}{\mathfrak{c}_2\mathfrak{c}_3}\mathfrak{P}_1\gamma^\nu\gamma_{\alpha}^\perp\gamma^\rho\gamma_\beta^\perp+
    \frac{\eta_\perp^{\alpha\beta}+\mathfrak{t}_3\hat{F}_\text{c}^{\alpha\beta}}{\mathfrak{c}_1\mathfrak{c}_2}\gamma_{\alpha}^\perp\gamma^\nu\gamma_\beta^\perp\gamma^\rho\mathfrak{P}_3+\frac{\eta_\perp^{\alpha\beta}+\mathfrak{t}_2\hat{F}_\text{c}^{\alpha\beta}}{\mathfrak{c}_1\mathfrak{c}_3}\gamma_{\alpha}^\perp\gamma^\nu\mathfrak{P}_2\gamma^\rho\gamma_\beta^\perp,\nonumber
    \\
    \mathfrak{T}_2^{\nu\rho}&:=\mathfrak{P}_1\gamma^\nu\gamma^\alpha_\perp\mathfrak{e}_2\gamma^\rho\gamma_{\alpha}^\perp\mathfrak{e}_3+\gamma^\alpha_\perp\mathfrak{e}_1\gamma^\nu\gamma_{\alpha}^\perp\mathfrak{e}_2\gamma^\rho\mathfrak{P}_3+\gamma^\alpha_\perp\mathfrak{e}_1\gamma^\nu\mathfrak{P}_2\gamma^\rho\gamma_{\alpha}^\perp\mathfrak{e}_3,\nonumber\\
    \mathfrak{T}_3^{\nu\rho}&:= \mathfrak{P}_1\gamma^\nu\mathfrak{P}_2 \gamma^\rho\mathfrak{P}_3
\end{align}
where 
\begin{align}
    \mathfrak{P}_1&:=\Pi_1\Bigg(
        \underbrace{-\frac{s_3k_\parallel+s_2p_{2\parallel}}{s_1+s_2+s_3}}_{a_1}+\underbrace{\frac{\mathfrak{t}_2\mathfrak{t}_3\hat{F}_\text{c}p_{1\perp}-\mathfrak{t}_2p_{2\perp}-\mathfrak{t}_3k_\perp}{\mathfrak{t}_1+\mathfrak{t}_2+\mathfrak{t}_3-\mathfrak{t}_1\mathfrak{t}_2\mathfrak{t}_3}}_{b_1}
    \Bigg)\nonumber\\
    \mathfrak{P}_2&:=\Pi_2\Bigg(\underbrace{\frac{s_1p_{2\parallel}-s_3p_{1\parallel}}{s_1+s_2+s_3}}_{a_2}+\underbrace{\frac{-\mathfrak{t}_1\mathfrak{t}_3\hat{F}_\text{c}k_{\perp}-\mathfrak{t}_3p_{1\perp}+\mathfrak{t}_1p_{2\perp}}{\mathfrak{t}_1+\mathfrak{t}_2+\mathfrak{t}_3-\mathfrak{t}_1\mathfrak{t}_2\mathfrak{t}_3}}_{b_2}\Bigg)\nonumber\\
    \mathfrak{P}_3&:=\Pi_3\Bigg(\underbrace{\frac{s_2p_{1\parallel}+s_1k_{\parallel}}{s_1+s_2+s_3}}_{a_3}+\underbrace{\frac{\mathfrak{t}_1\mathfrak{t}_2\hat{F}_\text{c}p_{2\perp}+\mathfrak{t}_1k_{\perp}+\mathfrak{t}_2p_{1\perp}}{\mathfrak{t}_1+\mathfrak{t}_2+\mathfrak{t}_3-\mathfrak{t}_1\mathfrak{t}_2\mathfrak{t}_3}}_{b_3}\Bigg)\nonumber.
\end{align}
Evaluation of these traces is performed using the \textsc{Mathematica} package \textsc{FeynCalc} \cite{Shtabovenko:2020gxv}. Due to the presence of $\gamma^5$ matrices, the results consist of contractions of external momenta with Levi-Civita tensors. These can be further reduced by using the symmetry condition that $\Gamma^{\mu\nu\rho}_{a\gamma\gamma}$ must be symmetric under the exchange of the $\nu$ and $\rho$ indices. As an example, we give the result of the shortest trace explicitly:
\begingroup
\allowdisplaybreaks
\begin{align}
    \text{Tr}\left[\gamma^5\gamma^\mu\mathfrak{T}_1^{\nu\rho}\right]&=4i\left[\mathfrak{t}_1\frac{\left(\eta_\perp^{\alpha\nu}\hat{\tilde{F}}_\text{c}^{\mu\rho}-\eta_\perp^{\nu\mu}\hat{\tilde{F}}_\text{c}^{\alpha\rho}\right)\left(b_{1,\alpha}-\mathfrak{c}_1a_{1,\alpha}\right)+a_{1,\alpha}\mathfrak{s}_1\hat{F}_\text{c}^{\mu\nu}\hat{\tilde{F}}_\text{c}^{\alpha\rho}
    }{\mathfrak{c}_2\mathfrak{c}_3}\right.\nonumber\\
    &\phantom{=4i\Bigg[}+\mathfrak{t}_2\frac{\left(\eta_\perp^{\alpha\nu}\hat{\tilde{F}}_\text{c}^{\mu\rho}+\eta_\perp^{\alpha\rho}\hat{\tilde{F}}_\text{c}^{\mu\nu}+\eta_\perp^{\nu\rho}\hat{\tilde{F}}_\text{c}^{\alpha\mu}\right)\left(b_{2,\alpha}-\mathfrak{c}_2a_{2,\alpha}\right)}{\mathfrak{c}_1\mathfrak{c}_3}\nonumber\\
    &\phantom{=4i\Bigg[}-\left.\mathfrak{t}_3\frac{\left(\eta_\perp^{\alpha\rho}\hat{\tilde{F}}_\text{c}^{\mu\nu}-\eta_\perp^{\mu\rho}\hat{\tilde{F}}_\text{c}^{\alpha\nu}\right)\left(b_{3,\alpha}+\mathfrak{c}_3a_{3,\alpha}\right)+a_{3,\alpha}\mathfrak{s}_3\hat{F}_\text{c}^{\mu\rho}\hat{\tilde{F}}_\text{c}^{\alpha\nu}
    }{\mathfrak{c}_1\mathfrak{c}_2}\right]\nonumber\\
    &\equiv \frac{1}{\mathfrak{c}_1\mathfrak{c}_2\mathfrak{c}_3}\sum\limits_{i=1}^3\mathfrak{s}_i\left(a_{i,\alpha}\left[\mathfrak{c}_iC_{1,i}^{\alpha\mu\nu\rho}+\mathfrak{s}_iC_{2,i}^{\alpha\mu\nu\rho}\right]+b_{i,\alpha}C_{3,i}^{\alpha\mu\nu\rho}\right).\label{eq:trace}
\end{align}
\endgroup
The above expression is a shorthand decomposition that we introduce to facilitate the integration. The individual tensors in Eq.~\eqref{eq:trace} correspond to lengthy linear combinations of products of Levi-Civita tensors and metric tensors, which do not depend on any remaining integration variables. Similar shorthand decompositions can be obtained for the remaining traces in Eq.~\eqref{eq:operators}, yielding
\begin{align}
    \text{Tr}\left[\gamma^5\gamma^\mu\mathfrak{T}_2^{\nu\rho}\right]&=\mathfrak{c}_1\mathfrak{c}_2\mathfrak{c}_3\sum\limits_{i=1}^3\left(a_{i,\alpha}\left[\mathfrak{t}_1C_{4,i}^{\alpha\mu\nu\rho}+\mathfrak{t}_2C_{5,i}^{\alpha\mu\nu\rho}+\mathfrak{t}_3C_{6,i}^{\alpha\mu\nu\rho}+\mathfrak{t}_1\mathfrak{t}_2C_{7,i}^{\alpha\mu\nu\rho}+\mathfrak{t}_1\mathfrak{t}_3C_{8,i}^{\alpha\mu\nu\rho}\right.\right.\nonumber\\
    &\phantom{=\mathfrak{c}_1\mathfrak{c}_2\mathfrak{c}_3\sum}\left.+\mathfrak{t}_2\mathfrak{t}_3C_{9,i}^{\alpha\mu\nu\rho}+\mathfrak{t}_1\mathfrak{t}_2\mathfrak{t}_3C_{10,i}^{\alpha\mu\nu\rho}\right]+\frac{b_{i,\alpha}}{\mathfrak{c}_i}\left[\mathfrak{t}_1C_{11,i}^{\alpha\mu\nu\rho}+\mathfrak{t}_2C_{12,i}^{\alpha\mu\nu\rho}+\mathfrak{t}_3C_{13,i}^{\alpha\mu\nu\rho}\right.\nonumber\\
    &\phantom{=\mathfrak{c}_1\mathfrak{c}_2\mathfrak{c}_3\sum}\left.\left.+\mathfrak{t}_1\mathfrak{t}_2C_{14,i}^{\alpha\mu\nu\rho}+\mathfrak{t}_1\mathfrak{t}_3C_{15,i}^{\alpha\mu\nu\rho}+\mathfrak{t}_2\mathfrak{t}_3C_{16,i}^{\alpha\mu\nu\rho}\right]\right),
\end{align}
and
\begin{align}
    \text{Tr}\left[\gamma^5\gamma^\mu\mathfrak{T}_3^{\nu\rho}\right]&=\mathfrak{c}_1\mathfrak{c}_2\mathfrak{c}_3\bigg(a_{1,\alpha}a_{2,\beta}a_{3,\tau}\left[D_{1}^{\alpha\beta\tau\mu\nu\rho}+\mathfrak{t}_1D_{2}^{\alpha\beta\tau\mu\nu\rho}+\mathfrak{t}_2D_{3}^{\alpha\beta\tau\mu\nu\rho}+\mathfrak{t}_3D_{4}^{\alpha\beta\tau\mu\nu\rho}\right.\nonumber\\
    &\phantom{=}\left.+\mathfrak{t}_1\mathfrak{t}_2D_{5}^{\alpha\beta\tau\mu\nu\rho}+\mathfrak{t}_1\mathfrak{t}_3D_{6}^{\alpha\beta\tau\mu\nu\rho}+\mathfrak{t}_2\mathfrak{t}_3D_{7}^{\alpha\beta\tau\mu\nu\rho}+\mathfrak{t}_1\mathfrak{t}_2\mathfrak{t}_3D_{8}^{\alpha\beta\tau\mu\nu\rho}\right]\nonumber\\
    &\phantom{=}+\frac{b_{1,\alpha}a_{2,\beta}a_{3,\tau}}{\mathfrak{c}_{1}}\left[D_{9}^{\alpha\beta\tau\mu\nu\rho}+\mathfrak{t}_{2}D_{10}^{\alpha\beta\tau\mu\nu\rho}+\mathfrak{t}_{3}D_{11}^{\alpha\beta\tau\mu\nu\rho}+\mathfrak{t}_{2}\mathfrak{t}_{3}D_{12}^{\alpha\beta\tau\mu\nu\rho}\right]\nonumber\\
    &\phantom{=}+\frac{a_{1,\alpha}b_{2,\beta}a_{3,\tau}}{\mathfrak{c}_{2}}\left[D_{13}^{\alpha\beta\tau\mu\nu\rho}+\mathfrak{t}_{1}D_{14}^{\alpha\beta\tau\mu\nu\rho}+\mathfrak{t}_{3}D_{15}^{\alpha\beta\tau\mu\nu\rho}+\mathfrak{t}_{1}\mathfrak{t}_{3}D_{16}^{\alpha\beta\tau\mu\nu\rho}\right]\nonumber\\
    &\phantom{=}+\frac{a_{1,\alpha}a_{2,\beta}b_{3,\tau}}{\mathfrak{c}_{3}}\left[D_{17}^{\alpha\beta\tau\mu\nu\rho}+\mathfrak{t}_{1}D_{18}^{\alpha\beta\tau\mu\nu\rho}+\mathfrak{t}_{2}D_{19}^{\alpha\beta\tau\mu\nu\rho}+\mathfrak{t}_{1}\mathfrak{t}_{2}D_{20}^{\alpha\beta\tau\mu\nu\rho}\right]\nonumber\\
    &\phantom{=}+\frac{b_{1,\alpha}b_{2,\beta}a_{3,\tau}}{\mathfrak{c}_{1}\mathfrak{c}_{2}}\left[D_{21}^{\alpha\beta\tau\mu\nu\rho}+\mathfrak{t}_{3}D_{22}^{\alpha\beta\tau\mu\nu\rho}\right]+\frac{b_{1,\alpha}a_{2,\beta}b_{3,\tau}}{\mathfrak{c}_{1}\mathfrak{c}_{3}}\left[D_{23}^{\alpha\beta\tau\mu\nu\rho}+\mathfrak{t}_{2}D_{24}^{\alpha\beta\tau\mu\nu\rho}\right]\nonumber\\
    &\phantom{=}+\frac{a_{1,\alpha}b_{2,\beta}b_{3,\tau}}{\mathfrak{c}_{2}\mathfrak{c}_{3}}\left[D_{25}^{\alpha\beta\tau\mu\nu\rho}+\mathfrak{t}_{1}D_{26}^{\alpha\beta\tau\mu\nu\rho}\right]+\frac{b_{1,\alpha}b_{2,\beta}b_{3,\tau}}{\mathfrak{c}_{1}\mathfrak{c}_{2}\mathfrak{c}_{3}}D_{27}^{\alpha\beta\tau\mu\nu\rho}\nonumber\\
    &\phantom{=}+m_f^2\sum\limits_{i=1}^3\left(a_{i,\alpha}\left[\mathfrak{t}_1C_{17,i}^{\alpha\mu\nu\rho}+\mathfrak{t}_2C_{18,i}^{\alpha\mu\nu\rho}+\mathfrak{t}_3C_{19,i}^{\alpha\mu\nu\rho}+\mathfrak{t}_1\mathfrak{t}_2C_{20,i}^{\alpha\mu\nu\rho}+\mathfrak{t}_1\mathfrak{t}_3C_{21,i}^{\alpha\mu\nu\rho}\right.\right.\nonumber\\
    &\phantom{=}+\mathfrak{t}_2\mathfrak{t}_3C_{22,i}^{\alpha\mu\nu\rho}+\mathfrak{t}_1\mathfrak{t}_2\mathfrak{t}_3C_{23,i}^{\alpha\mu\nu\rho}+\frac{b_{i,\alpha}}{\mathfrak{c}_i}\left[\mathfrak{t}_1C_{24,i}^{\alpha\mu\nu\rho}+\mathfrak{t}_2C_{25,i}^{\alpha\mu\nu\rho}+\mathfrak{t}_3C_{26,i}^{\alpha\mu\nu\rho}\right.\nonumber\\
    &\phantom{=}\left.\left.+\mathfrak{t}_1\mathfrak{t}_2C_{27,i}^{\alpha\mu\nu\rho}+\mathfrak{t}_1\mathfrak{t}_3C_{28,i}^{\alpha\mu\nu\rho}+\mathfrak{t}_2\mathfrak{t}_3C_{29,i}^{\alpha\mu\nu\rho}\right]\right)\bigg).
\end{align}\label{eq:traces}
Explicit expressions for the Bose-symmetrized versions of the tensors \(C_{i,j}^{\alpha\mu\nu\rho}\) and \(D_{i}^{\alpha\beta\tau\mu\nu\rho}\) can be found in the \textsc{Mathematica} notebook file located in the aforementioned \textsc{GitHub} page. Combined with Eq.~\eqref{eq:operators}, we see that parts of the integrand that depend on the proper time variable consist of powers of trigonometric functions and a complicated exponential function.

\subsection{Proper Time Integrals}\label{sec:proper_time_integrals}
\begin{figure}[!htb]
    \centering
    \includegraphics[width=0.65\textwidth]{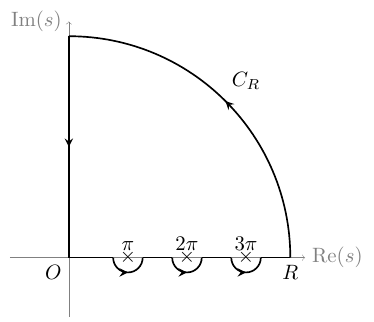}
    \caption{One of the many possible contours of integration for the proper-time integral given in Eq.~\eqref{eq:vertex_loop_final_expanded}, for a given subdomain of the $v_1$ and $v_2$ variables. The orientation of the path is determined dynamically using the exclusion functions defined in Eqs.~\eqref{eq:exclusion_functions}.}
    \label{fig:contour}
\end{figure}

Our goal is to provide an exact solution that can be numerically computed for all combinations of perpendicular momenta and magnetic field strengths, while allowing suitable approximations in different regimes. We begin by changing the integration variables as 
\[
\beta s_1\to s\,v_1,\quad \beta s_2\to s\,v_2,\quad \text{and} \quad \beta s_3\to s\,(1-v_1-v_2)\equiv s\,v_3,
\]
with the integration domain given by
\begin{equation}
    (s,v_1,v_2)\in\mathbb{R}_{\geq0}\times\{(v_1,v_2)~|~0\leq v_1\leq1,\,0\leq v_2\leq 1-v_1\}.
\end{equation}
We first integrate over the $s$ variable by treating it as a complex number, and then perform the remaining $v_i$ integrals numerically. By expanding the trace result obtained above, we obtain proper time integrals of the form 
\begin{equation*}
    K\int\text{d}s~\underbrace{\csc^n(s)T_0 s^m e^{\frac{i}{\beta}\csc(s)\left[p_{1\perp}^2T_1+p_{2\perp}^2T_2+k_{\perp}^2T_3+2 k_{\perp}\hat{F}_\text{c}p_{1\perp}T_4\right]}e^{i s P(v_1,v_2)}}_{=:J_{n,m,T_0}(s,v_1,v_2)},
\end{equation*}
with
\begin{equation}
    P(v_1,v_2)=\frac{1}{\beta }\left[v_1v_3k_\parallel^2+v_2v_3p_{1\parallel}^2+v_1v_2p_{2\parallel}^2-m_f^2\right],
\end{equation}
where $K$ is independent of $s$, the $T_i$ are products of sine and cosine functions of $s$, $v_1$, and $v_2$, and $n,m\in\mathbb{Z}$. Essential singularities from the complicated exponential in Eq.~\eqref{eq:operators}, as well as poles from the trace term, are all located on the real axis at values $s\in\mathbb{Z}^\ast\pi$. Since we are only interested in the positive real axis, we use an integration path resembling a quarter circle, as illustrated in Figure~\ref{fig:contour}.

To be rigorous, we fix the external momenta and use a path-valued function of the $v_1$ and $v_2$ variables, since the behavior of the quarter circle for $R\to\infty$ and $\varepsilon\to0$ can vary over different subdomains and external momenta. We decompose a general path as 
\begin{equation*}
    C_R=C_\text{Arc}\cup C_\text{Im}\cup C_{\text{Re},\varepsilon}\bigcup\limits_{r=1}^{\floor*{\frac{R}{\pi}}} C_{\varepsilon,r},
\end{equation*} 
where $C_\text{Arc}$ is the quarter-circle with radius $R$, $C_\text{Im}$ is the path along the imaginary axis, $C_{\text{Re},\varepsilon}$ is the path along the real axis with periodic gaps of diameter $2\varepsilon$ around the poles, and for each $r\in\{1,\ldots,\floor*{\frac{R}{\pi}}\}$, $C_{\varepsilon,r}$ is a semicircle of radius $\varepsilon$ circumventing the $r$-th pole. (We do not assume any particular orientation for these paths at this stage.)

We first determine the orientation of $C_\text{Arc}$ by setting $s\to Re^{i\theta}$ with $\theta\in [0,\frac{\pi}{2}]$ (first quadrant) or $\theta\in[-\frac{\pi}{2},0]$ (fourth quadrant). In this case, the magnitude of each possible trigonometric product is bounded by
\begin{align}
    \left|\csc^n[Re^{i\theta}]T_i\right|&\leq \frac{\cosh[k v_1 R \sin(\theta)]\cosh[k v_2 R \sin(\theta)]\cosh[k v_3 R \sin(\theta)]}{|\sinh^n(R \sin(\theta))|}\nonumber\\
    &\leq \frac{e^{k R \sin(\theta)}}{|\sinh^n(R \sin(\theta))|}\nonumber\\
    &\leq 2^n\frac{e^{nR\sin(\theta)}}{\left|e^{R\sin(\theta)}-e^{-R\sin(\theta)}\right|^n}\xrightarrow{R\to\infty}2^n,
\end{align}
where $0<k\leq n\leq4$. This bound is independent of $\theta$ and $k$. Consequently, the exponential function with trigonometric arguments is also bounded in this limit. On the other hand, we obtain a polynomial in $s$ of order $m$, resulting in an overall divergence of order $R^{m+1}$ (the additional linear divergence arises from the change of variables, $\text{d}s=i R e^{i\theta}\text{d}\theta$). The remaining non-trigonometric exponential must vanish faster than $R^{-(m+1)}$. In fact,
\begin{equation}
    \left|e^{iRe^{i\theta}P(v_1,v_2)}\right|=e^{-R\sin(\theta)P(v_1,v_2)},
\end{equation}
which vanishes exponentially as $R\to\infty$, provided that the overall sign of the exponent is negative. This condition depends on the external momenta, the fermion mass, and the values of $v_i$, which in turn determine the path orientation.

A similar discussion applies to the behavior along $C_{\varepsilon,r}$. Setting $s\to r\pi+\varepsilon e^{i\theta}$ with $\theta\in[0,\pi]$ (upper semicircle) or $\theta\in[-\pi,0]$ (lower semicircle), the magnitude of the term with the linear exponent converges to one as $\varepsilon\to0$. However, the trigonometric functions, specifically $\csc^n$, diverge in this limit because the semicircles are centered around the poles. Expanding $\csc^n(s)$ around $s=r\pi$ for very small $\varepsilon$, we find
\begin{align}
    \left|\csc^n[r\pi+\varepsilon e^{i\theta}]T_i\right|&\approx \frac{|T_i|_{s=r\pi}}{\varepsilon^n}.
\end{align} 
Combined with the extra factor of $\varepsilon$ from the change of variables, we require the remaining exponential function to vanish faster than $\varepsilon^{n-1}$. Explicitly,
\begin{align}
    \left|e^{\frac{i}{\beta }\csc[Re^{i\theta}]\left[p_{1\perp}^2T_1+p_{2\perp}^2T_2+k_{\perp}^2T_3+2 k_{\perp}\hat{F}_\text{c}p_{1\perp}T_4\right]}\right|&\approx e^{\frac{(-1)^r}{\beta }\frac{\sin(\theta)}{\varepsilon}\left[p_{1\perp}^2T_1+p_{2\perp}^2T_2+k_{\perp}^2T_3+2 k_{\perp}\hat{F}_\text{c}p_{1\perp}T_4\right]_{s=r\pi}},
\end{align}
which vanishes exponentially as $\varepsilon\to0$, depending on the sign of the exponent. This result is crucial, as it indicates that some of the poles must be dynamically excluded.

To construct a compact general result, we define the exclusion functions 
\begin{align}
    f_\text{Arc}&:=\text{sgn}\left(P(v_1,v_2)\right),\\[1mm]
    f_{r}&:=\frac{1+f_\text{Arc}\cdot\text{sgn}\left(\frac{(-1)^r}{e_f}\left[p_{1\perp}^2T_1+p_{2\perp}^2T_2+k_{\perp}^2T_3+2 k_{\perp}\hat{F}_\text{c}p_{1\perp}T_4\right]_{s=r\pi} \right)}{2}.\label{eq:exclusion_functions}
\end{align}
Using these functions, the general result of the proper-time integral for a specific $J_{n,T_0}$ can be written as
\begin{equation}
    \int\limits_0^\infty\text{d}s~J_{n,m,T_0}(s,v_1,v_2)=i\int\limits_0^{f_\text{Arc}\cdot\infty}\text{d}s~J_{n,m,T_0}(i s,v_1,v_2)+2\pi i\sum\limits_{r=1}^\infty f_r  \mathop{\mathrm{Res}}_{s=r\pi}\left[J_{n,m,T_0}\right].\label{eq:final_proper_time}
\end{equation}
Since there are no poles along the imaginary axis, the first integral on the right-hand side of Eq.~\eqref{eq:final_proper_time} can be evaluated numerically without difficulty. As an example for the second term, we compute the residues for the integral corresponding to the tensor $C_{1,1}^{\alpha\mu\nu\rho}$, i.e., the case where 
\begin{equation}
    \csc^n(s)T_0=\frac{\mathfrak{s}_1\mathfrak{c}_1}{\left[(\mathfrak{c}_1\mathfrak{c}_2\mathfrak{c}_3)(\mathfrak{t}_1+\mathfrak{t}_2+\mathfrak{t}_3-\mathfrak{t}_1\mathfrak{t}_2\mathfrak{t}_3)\right]^2}=\csc^2(s)\frac{\sin(2s v_1)}{2},
\end{equation}
with $n=2$ and $m=1$. Due to the $\csc$ functions present, the Laurent series around the poles has a convergence radius of $\pi$, meaning that we must expand the series around each pole individually. Using the multinomial theorem combined with the generalized Cauchy product, we find the coefficients for the exponential term. The result can be summarized as 
\begin{align}
    \mathop{\mathrm{Res}}_{s=r\pi}\left[J_{1,1,T_0}\right]&=\sum\limits_{l=0}^\infty\sum\limits_{k=0}^{l+1}\sum\limits_{j=0}^k\frac{i^l(-1)^{\frac{j}{2}+1+\floor*{\frac{k-j}{2}}}2^j(j-1)B_j}{l!j!(k-j)!}a_{k-j}(2v_1)\nonumber\\
    &\times\left(r\pi c_{l-k+1}+\sum\limits_{m=0}^{l-k}c_m\frac{(iP(v_1,v_2)r\pi)^{l-k-m}e^{ir\pi P(v_1,v_2)}[1+l-k-m+iP(v_1,v_2)]}{(2-k+l-m)!}\right),\label{eq:residue}
\end{align}\label{eq:result}
with
\begingroup
\allowdisplaybreaks
\begin{align}
    a_n(\omega)&:=\sin(r\pi \omega)^{\frac{1+(-1)^{n}}{2}}\cos(r\pi \omega)^{\frac{1-(-1)^{n}}{2}}\omega^n,\\[1mm]
    b_n(\omega)&:=[-\sin(r\pi \omega)]^{\frac{1-(-1)^{n}}{2}}\cos(r\pi \omega)^{\frac{1+(-1)^{n}}{2}}\omega^n,\\[1mm]
    c_0&:=d_0^l,\\[1mm]
    c_m&:=\frac{1}{m d_0}\sum\limits_{n=1}^m(nl-m+n)c_md_{m-n},\\[1mm]
    d_m&:=\sum\limits_{n=0}^m\frac{B_ni^{n}(-1)^{r+\floor*{\frac{m-n}{2}}}(1-2^{n-1})}{n!(m-n)!2}\bigg[2p_{1\perp}^2b_{m-n}(2v_1+2v_2-1)-2k_\perp^2b_{m-n}(1)\nonumber\\
    &\phantom{=\sum}+k_\perp\hat{F}_\text{c}p_{1\perp}\big[a_{m-n}(2v_1+2v_2-1)+a_{m-n}(1-2v_2)-a_{m-n}(2v_1-1)-a_{m-n}(1)\big]\nonumber\\
    &\phantom{=\sum}+2p_{2\perp}^2b_{m-n}(2v_1-1)\bigg].
\end{align}
\endgroup
Here, $B_j$ is the $j$-th Bernoulli number. Although this expression may seem tedious—especially given that there are 57 other tensors to evaluate—it is not problematic because the sequences $a_n$, $b_n$, $c_n$, and $d_n$ are identical across all tensors, with the only differences arising from the first term in Eq.~\eqref{eq:residue}. Furthermore, since these coefficients are reused repeatedly, one can employ \textit{memoization} techniques to speed up the computation. Although the summation above does not have a closed form, one can use as many terms in the expression as needed to achieve the desired precision. The speed of convergence increases proportionally to the strength of the magnetic field. Furthermore, considering the conservation of momentum at the vertex, the exclusion function $f_r$ turns out to be independent on the order of the pole, $r$, meaning either all poles are included in the integration for a given $(v_1,v_2)$ pair, or none of them are. This allows us to sum over all poles at once, and numerically integrate the acquired closed form afterwards. For example, assuming $l=0$, the residue from the Eq.~\eqref{eq:residue} can be summed as
\begin{align}
    \sum\limits_{r=1}^\infty\mathop{\mathrm{Res}}_{s=r\pi}\left[J_{1,1,T_0}\right]&=-\frac {-2 \pi  \sin (2 \pi  v_1) P (v_1, 
    v_2) \sin (\pi  P (v_1, v_2)) + 
 4 \pi  v_1 + \sin (4 \pi  v_1)} {8 [\cos (\pi  P (v_1, 
        v_2)) - \cos (2
           \pi  v_1)]^2}\nonumber\\
           &-\frac{2 (\sin (2 \pi  v_1) + 2 \pi  v_1 \cos
           (2 \pi  v_1)) \cos (\pi  P (v_1, v_2)) }{8 [\cos (\pi  P (v_1, 
           v_2)) - \cos (2
              \pi  v_1)]^2}
\end{align}
Closed form expressions for the remaining tensors ($l=0$) can also be found in the \textsc{Mathematica} notebook in the \textsc{GitHub} page.
\section{Special Cases}\label{sec:special_cases}
In this section, we briefly discuss the special cases of the vertex tensor $\Gamma_{a\gamma\gamma}^{\mu\nu\rho}$ that can be obtained from the general result derived in the previous section. These special cases are particularly relevant for practical applications, and useful for discussing the validity of the proper time integrals calculated above. In particular, we consider the cases where the external magnetic field is either very strong, or non-existent. Also, we will probe our original result in the interesting case of vanishing perpendicular momenta.
\subsection{Vanishing Magnetic Field}
The result obtained in the end of Section~\ref{sec:proper_time_integrals} looks linearly divergent in the limit of vanishing magnetic field, $\beta\to0$. This is expected, since the substitution $\beta s_i\to sv_i$ is undefined for $\beta=0$, and the limit is not interchangeable with the integration. However, we can still assign a finite value for the vanishing case by taking the limit before performing the substitution. 
As for the tensorial structure, imposing the momentum conservation condition, the Bose-symmetric pseudotensor of third order without external magnetic fields can be written as\cite{Bell:348417}
\begin{align}
    \Gamma^{\mu\nu\rho}_{\emptyset}&=c^\emptyset_1 \varepsilon^{\nu\rho\alpha\beta}p_{1\alpha} p_{2\beta}(p_{1}^{\mu}+p_{2}^{\mu})+c^\emptyset_2 p_{1\alpha} p_{2\beta}(\varepsilon^{\mu\nu\alpha\beta}p_{2}^\rho-\varepsilon^{\mu\rho\alpha\beta}p_{1}^\nu)\nonumber\\&+c^\emptyset_3 \left[p_{1\alpha} p_{2\beta}(\varepsilon^{\mu\nu\alpha\beta}p_{1}^\rho-\varepsilon^{\mu\rho\alpha\beta}p_{2}^\nu)+(p\cdot k)\varepsilon^{\mu\nu\rho\alpha}(p_{1\alpha}-p_{2\alpha})\right].
\end{align}
For this reason, we resume the calculation from our intermediate result in Eq.~\eqref{eq:operators}, and take the limit $\beta\to0$. We obtain
\begin{equation}
    \Gamma^{\mu\nu\rho}_{\emptyset}=\frac{\pi^2}{2}\sum\limits_f\frac{e_f^2C_{ff}}{\Lambda}\iiint\text{d}s_1~\text{d}s_2~\text{d}s_3~\frac{\text{Tr}\left[\gamma^5\gamma^\mu\mathfrak{T}_\emptyset^{\nu\rho}\right]}{(s_1+s_2+s_3)^2}e^{i\frac{p_{1}^2s_2s_3+p_{2}^2s_1s_2+k^2s_1s_3}{s_1+s_2+s_3}-im_f^2(s_1+s_2+s_3)}\label{eq:vertex_loop_zero0}
\end{equation}
where
\begin{equation}
    \mathfrak{T}_\emptyset^{\nu\rho}:=\left(m_f-\frac{s_3\slashed{k}+s_2\slashed{p}_2}{s_1+s_2+s_3}\right)\gamma^\nu\left(m_f+\frac{s_1\slashed{p}_2-s_3\slashed{p}_1}{s_1+s_2+s_3}\right)\gamma^\rho\left(m_f+\frac{s_2\slashed{p}_1+s_1\slashed{k}}{s_1+s_2+s_3}\right)
\end{equation}
We have omitted the $\delta^{\left(4\right)}(k-p_{2}-p_{1})$ term from the vertex tensor for brevity, but it is always implied.
Since no magnetic field is present, we now perform the slightly different substitution 
\begin{equation}
    s_1\to s\,v_1,\quad s_2\to s\,(1-v_1-v_3) \equiv s\,v_2,\quad \text{and} \quad s_3\to s\,v_3,
\end{equation}
with the same integration domain as in Section~\ref{sec:proper_time_integrals}, where we use $v_3$ instead of $v_2$ for later convenience. As a result of this substitution, the integrand obtains an additional factor of $s^2$, removing the divergence. We have
\begin{equation}
    \Gamma^{\mu\nu\rho}_{\emptyset}=\frac{\pi^2}{2}\sum\limits_f\frac{e_f^2C_{ff}}{\Lambda}\iiint\text{d}v_1~\text{d}v_3~\text{d}s~\text{Tr}\left[\gamma^5\gamma^\mu\mathfrak{T}_\emptyset^{\nu\rho}\right]\exp\bigg[is(\underbrace{p_{1}^2v_2v_3+p_{2}^2v_1v_2+k^2v_1v_3-m_f^2}_{=:K})\bigg]\label{eq:vertex_loop_zero1}
\end{equation}
where the only $s$ dependence is in the exponential term. Therefore, the integral over the proper time variable can be identified as a Fourier transform of the Heaviside function $\theta(s)$ in distributional sense, which gives us 
\begin{equation}
    \Gamma^{\mu\nu\rho}_{\emptyset}=\frac{\pi^2}{2}\sum\limits_f\frac{e_f^2C_{ff}}{\Lambda}\iint\text{d}v_1~\text{d}v_2~\text{Tr}\left[\gamma^5\gamma^\mu\mathfrak{T}_\emptyset^{\nu\rho}\right]\left[\pi\delta(K)+i\frac{1}{\mathcal{P}(K)}\right].
\end{equation}\label{eq:vertex_loop_zero2}
To facilitate comparison with existing literature, we consider the on-shell limit for the photons, $p_1^2 = p_2^2 = 0$. In this case, the equation $K = 0$ admits no solutions for $v_1$ or $v_3$ within the integration domain, causing the delta function term to vanish. The remaining term can then be integrated in the standard way, replacing $\frac{1}{\mathcal{P}(K)}$ with $\frac{1}{K}$. After evaluating the trace, integration over the $v_1$ and $v_3$ variables is straightforward. Combined with the swapped external photons, we obtain the final form factors
\begin{align}
    c_1^\emptyset&=\frac{4\pi^2}{(k^2)^2}\sum\limits_f\frac{e_f^2C_{ff}}{\Lambda}\left(\frac{7k^2+48m_f^2}{3}-2 A+i(k^2+8m_f^2)B\right)\nonumber\\
    c_2^\emptyset&=\frac{4\pi^2}{(k^2)^2}\sum\limits_f\frac{e_f^2C_{ff}}{\Lambda}\left(\frac{10k^2+48m_f^2}{3}-4 A+i(k^2+8m_f^2)B\right)\nonumber\\
    c_3^\emptyset&=\frac{4\pi^2}{(k^2)^2}\sum\limits_f\frac{e_f^2C_{ff}}{\Lambda}\left(\frac{k^2+48m_f^2}{3}-2 A+8im_f^2B\right)
\end{align}
with
\begin{align}
    A&=m_f^2\left[\text{Li}_2\left(\frac{2}{1+i\sqrt{\frac{4m_f^2}{k^2}-1}}\right)+\text{Li}_2\left(\frac{2}{1-i\sqrt{\frac{4m_f^2}{k^2}-1}}\right)\right],\\
    B&=\sqrt{\frac{4m_f^2}{k^2}-1}\log\left(\frac{\sqrt{\frac{4m_f^2}{k^2}-1}+i}{\sqrt{\frac{4m_f^2}{k^2}-1}-i}\right).
\end{align}
Depending on the ratio $\frac{4m_f^2}{k^2}$, the dilogarithm functions can be expressed in terms of $\text{sin}^{-1}$ or complex logarithms. Our results thus agrees with the field-free effective couplings derived in the literature \cite{Bauer_2017}.
\subsection{Strong Magnetic Field}
Assuming a very strong magnetic filed from the start, i.e. $B\gg B_c$, we can actually use a completely different approach to calculate the vertex tensor. In fact, instead of using the Schwinger proper-time method, we can expand the momentum part of the field-dressed fermion propagator given in Eq.~\eqref{eq:fermion_propagator_fourier}~in terms of the Landau levels \cite{PhysRevD102114038}:
\begin{equation}
    S_f(p)=e^{-\frac{p_\perp^2}{\beta}}\sum\limits_{n=0}^\infty\frac{(-1)^n D_n(p)}{p_\parallel^2-m_f^2-2n\beta+i\varepsilon},
\end{equation}
with 
\begin{align}
    D_n(p)&=(\slashed{p}_\parallel+m_f)(1+i\gamma_1\gamma_2\text{sgn}[\beta])L^0_n\left(\frac{2p_\perp^2}{\beta}\right)-(\slashed{p}_\parallel+m_f)(1-i\gamma_1\gamma_2\text{sgn}[\beta])L^0_{n-1}\left(\frac{2p_\perp^2}{\beta}\right)\nonumber\\&+4p_\perp^2L^1_{n-1}\left(\frac{2p_\perp^2}{\beta}\right),
\end{align}
where $L^k_n(x)$ are the Laguerre polynomials. This expansion has the advantage that for the magnetic field strength regime we are interested in, we can approximate the propagator by truncating the series at $n=0$, called the leading Landau level (LLL). Specifically, we have
\begin{equation}
    S_f(p)\approx ie^{-\frac{p_\perp^2}{\beta}}\underbrace{\frac{\slashed{p}_\parallel+m_f}{p_\parallel^2-m_f^2+i\varepsilon}\left(1+\gamma_1\gamma_2\text{sgn}[\beta]\right)}_{=:T_f(p)},
\end{equation}
which is much simpler to work with than with the full propagator. Note that the extra phase factor from the previous section is still present, and the integration regarding the spacetime coordinates is independent of the momentum part of the propagator. We can therefore resume the calculation from Eq.~\eqref{eq:vertex_loop_final}.\\
Specifically, we consider the case where the external photons are aligned with the external field, and their center of mass energy is equal to the ALP mass. which is assumed to be less then the Schwinger limit $m_a<2m_e$. This is motivated by the fact that photons with certain physical polarizations are expected to acquire a \textit{quasi mass} if a non-vanishing perpendicular momentum is present. Furthermore, this case is particularly interesting for our original result, since one needs an alternative approach for the proper time integrals, as explained in the Section~\ref{sec:discussion}. A significant advantage of this approach is that the end result is valid for \textit{all} magnetic field strengths and is not limited by computational resources. \\
With that in mind, the effective vertex tensor can be written as
\begin{align}
    \Gamma_{a\gamma\gamma}^{\mu\nu\rho}&=-\frac{2}{\left(2\pi \right)^2}\sum\limits_f\frac{e_f^2C_{ff}}{\beta^2\Lambda}
    \iiint\text{d}^2q_\perp~\text{d}^2\ell_\perp~\text{d}^4w~e^{i[q_\perp-w_\perp]\frac{2\hat{F}_\text{c}}{\beta }[w_\perp-\ell_\perp]-\frac{q_\perp^2+\ell_\perp^2+w_\perp^2}{\beta}}\\
    &\times\text{Tr}\left[\gamma^5\gamma^\mu T_f(k) \gamma^\nu T_f(p_1)\gamma^\rho T_f(0)\right],\nonumber
\end{align}
which, in terms of the perpendicular integration variables, has a similar structure to the full result. As such we can treat it in the same way as before, i.e. as Gaussian integrals. The calculation is straightforward, and we conveniently obtain
\begin{equation}
    \Gamma_{a\gamma\gamma}^{\mu\nu\rho}=i\pi\sum\limits_{f}\beta\frac{e_f^2C_{ff}}{\Lambda}\int \text{d}^2w_\parallel~\frac{G^{\mu\nu\rho}_1+G^{\mu\nu\rho\alpha}_2 w_{\parallel,\alpha}+G^{\mu\nu\rho\alpha\beta}_3 w_{\parallel,\alpha}w_{\parallel,\beta}+G^{\mu\nu\rho\alpha\beta\tau}_4 w_{\parallel,\alpha}w_{\parallel,\beta}w_{\parallel,\tau}}{[(w_\parallel-k_\parallel)^2-m_f^2+i\varepsilon][(w_\parallel-p_{1\parallel})^2-m_f^2+i\varepsilon][w_\parallel^2-m_f^2+i\varepsilon]}\label{eq:strong_inter},
\end{equation}
where $G_i$ represent the tensorial expressions obtained from the traces, with their explicit forms provided in the in the supplementary \textsc{GitHub} repository. This result is remarkably similar to integrals one would encounter in the context of usual fermionic loop calculations, with the only difference here being the dimensionality of the integration variables, $D=2$.\\
Before proceeding with the integration, we can already extract crucial information from the intermediate result above. Since the integral is independent of the magnitude of the external magnetic field in this limit, we can conclude that, with parallel external particles, observables proportional to the vertex tensor exhibit a quadratic dependence on the magnetic field strength.
The remaining integration can be performed in various, and from the four dimensional case well-known, ways. For each possible integral, their results are provided in the Appendix \ref{app:strong_mag}. Given the assumptions made above, after the addition of the Bose-symmetric counterpart, the magnetic addition of the absolute square of the effective coupling in the context of the kinematically constrained differential decay rate can be expressed as 
\begin{align}
    \left|g_{a\gamma\gamma,\parallel}^\text{eff.}\right|^2&=\frac{\pi^4}{4\Lambda^2m_a^4}\sum\limits_{f,f^\prime}\frac{\beta\beta^\prime e_f^2e_{f^\prime}^2C_{ff}C_{f^\prime f^\prime}}{m_f^2m_{f^\prime}^2\sqrt{\frac{4 m_f^2}{m_a^2}-1}\sqrt{\frac{4 m_{f^\prime}^2}{m_a^2}-1}} \Bigg\{m_a^4\sqrt{\frac{4 m_f^2}{m_a^2}-1}\sqrt{\frac{4 m_{f^\prime}^2}{m_a^2}-1}\nonumber\\
    &+64m_f^2 m_{f^\prime}^2\tan
    ^{-1}\left(\frac{m_a}{\sqrt{\frac{4 m_f^2}{m_a^2}-1}}\right)\tan
    ^{-1}\left(\frac{m_a}{\sqrt{\frac{4 m_{f^\prime}^2}{m_a^2}-1}}\right)\nonumber\\
    &\left.-4m_a^4\left[m_f^2\sqrt{\frac{4 m_{f^\prime}^2}{m_a^2}-1}\tan
    ^{-1}\left(\frac{m_a}{\sqrt{\frac{4 m_f^2}{m_a^2}-1}}\right)+m_{f^\prime}^2\sqrt{\frac{4 m_{f}^2}{m_a^2}-1}\tan
    ^{-1}\left(\frac{m_a}{\sqrt{\frac{4 m_{f^\prime}^2}{m_a^2}-1}}\right)\right]\right\}\label{eq:strong_mag_coupling}.
\end{align}
Note that this result is valid for ALP masses below the Schwinger limit. Furthermore, although all fermions are considered here, the analytical structure indicates that the contributions from lighter fermions, especially electrons, dominate the field-dressed vertex. As expected from the intermediate result above, the effective coupling grows linearly with the magnetic field strength
\section{Discussion}\label{sec:discussion}
We now briefly discuss the implications of the results derived in the preceding sections. The calculated field-dressed vertex tensor, $\Gamma_{a\gamma\gamma}^{\mu\nu\rho}$, represents the core quantity encapsulating the quantum corrections to the ALP-photon interaction within the external magnetic field. This tensor serves as the fundamental input for computing physical observables, such as the decay rate of ALPs into photons or the probability of photon-ALP conversion, within an effective Lagrangian framework. The magnitude and structure of these observables are inherently sensitive to the underlying model parameters, specifically the fermion-specific Wilson coefficients $C_{ff}$, the intrinsic vacuum ALP-photon coupling $g_{a\gamma\gamma}$, and the effective scale $\Lambda$, as indicated by Eq.~\eqref{eq:dim5operators}.
To gain concrete insight into the behavior of the vertex tensor $\Gamma_{a\gamma\gamma}^{\mu\nu\rho}$, we examine its properties under specific kinematic conditions and field strengths. While the general Lorentz decomposition in Eq.~\eqref{eq:vertex_decomposition} is extensive, it simplifies considerably in certain limits. For instance, taking photons on-shell ($p_1^2=p_2^2=0$) causes exactly half of the coefficients $c_i$ to vanish, reducing the complexity of the tensor structure.
To illustrate the impact of external magnetic fields, such as those encountered in astrophysical environments like magnetars or on laboratory experiments, we evaluate the remaining coefficients under specific, representative conditions. The most general form of our result is an infinite sum without a closed form, and the rate of convergence is directly proportional to the strength of the magnetic field. There are, however, some cases where the vertex tensor can be expressed in a closed form, making it valid for all magnetic field strengths. In order to demonstrate the qualitative behavior of the vertex tensor, and compare it to the approximate results from the previous section, we consider one such case, where the external particles are all parallel to the external magnetic field.\\
This case is analytically interesting and practically relevant, especially for helioscope experiments. Moreover, as mentioned in the previous section, the presence of a perpendicular momentum component is expected to induce a non-trivial dispersion relation for certain photon polarizations when the magnetic field is sufficiently strong, $B \gg B_c$ \cite{shabad2004photon}. This makes direct comparison with the LLL approach particularly challenging.\\
Analytically, in the on-shell case a vanishing perpendicular momentum removes the lengthy trigonometric exponential in Eq.~\eqref{eq:operators}, and the remaining integral can then be evaluated in a more direct manner. Besides the standard residue ansatz, one may also shift the proper-time variable infinitesimally along the imaginary axis and reduce the integral to a trivial gamma-function form. This was not feasible in the fully general case because of the trigonometric functions in the exponent.\\
\begin{figure}[!htb]
    \centering
    \includegraphics[width=0.9\textwidth]{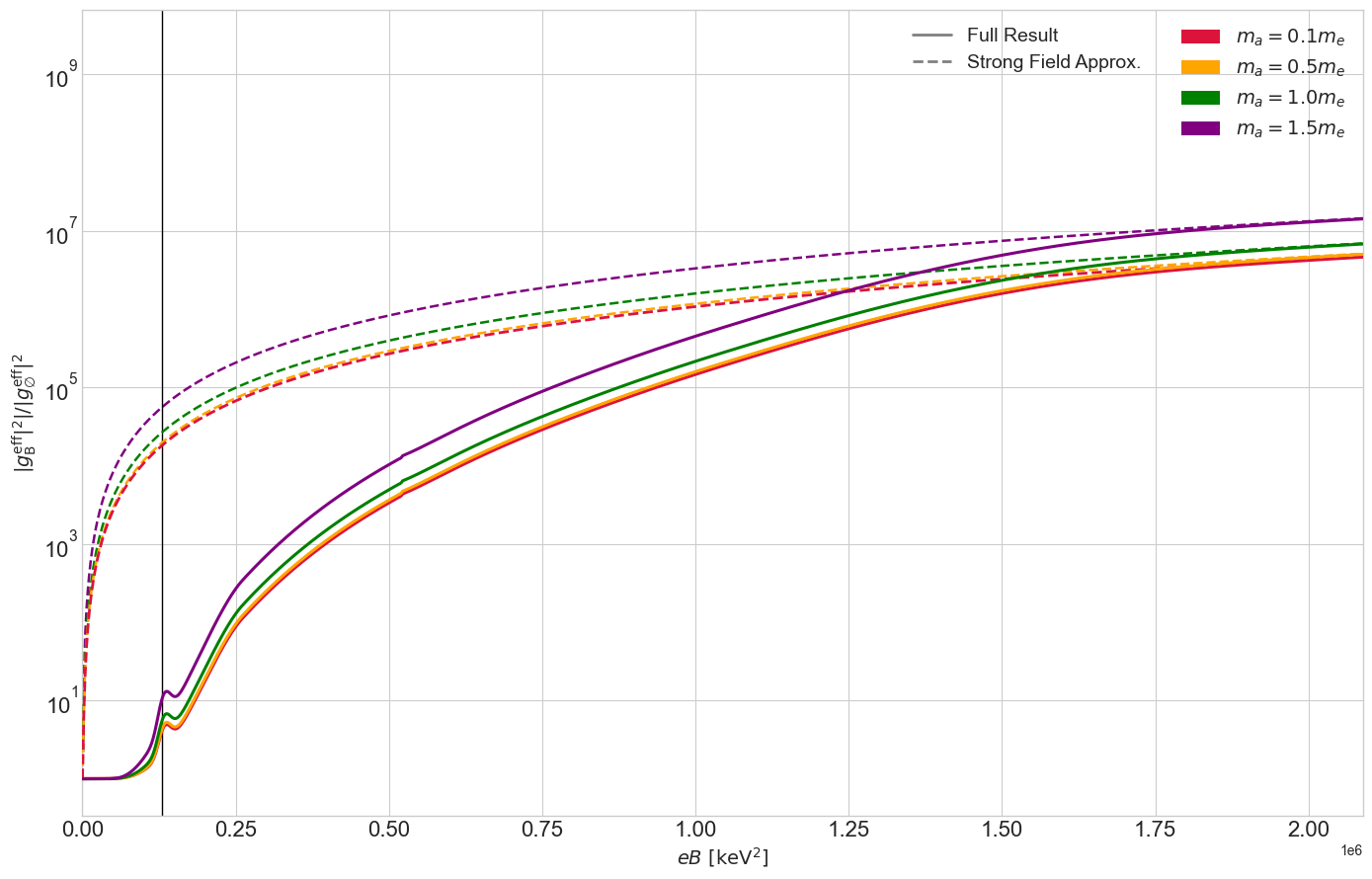}
    \caption{Ratio of the field-dressed ALP–photon coupling squared, $|g_B^\text{eff}|^2$, to the vacuum coupling squared, $|g_\emptyset^\text{eff}|^2$, as a function of the magnetic field strength, $\beta=eB$, for different ALP masses $m_a$. Solid lines: exact result from Eq.~\eqref{eq:operators}. Dashed lines: strong-field (LLL) approximation. Only ALP–electron coupling is included, with $\frac{g_{a\gamma\gamma}}{g_{aee}}=0.1$. The blue vertical line marks $eB=\frac{m_e^2}{2}$.}\label{fig:strong_field}
\end{figure}
After performing the proper-time integral, the remaining $v_i$-dependent terms are evaluated numerically, since most contributions reduce to special polylogarithms and logarithmic gamma integrals without known closed-form expressions.\\
As a demonstration of this kinematic configuration, we allow only ALP–electron interactions and set the ratio of the photonic to electron coupling to $0.1$. The results, in the context of the kinematically constrained differential decay rate, are shown in Figure~\ref{fig:strong_field} for several ALP masses $m_a<2m_e$. Above this mass range, additional decay channels open, introducing further complications. The solid lines show the exact result from Eq.~\eqref{eq:operators}, while the dashed lines correspond to the strong-field (LLL) approximation discussed in Section~\ref{sec:special_cases}. As expected, for lower field values where the strong-field approximation is invalid, the effective coupling changes little with $B$ until $eB\approx \frac{m^2}{2}$, where it rises sharply. Beyond this point, the effective coupling grows rapidly and approaches the strong-field limit. This enhancement is even more pronounced for ALP masses closer to the Schwinger threshold, $m_a \to 2m_e$.
By allowing non-zero perpendicular momenta, as explained above, the vertex term gets an non-closed form, with the speed of convergence being proportional to the magnetic field strength, making a direct comparision with the strong field approximation considerably more challanging. Furthermore, the behavior of the observables are also affected by the vacuum birefringence and cyclotron resonance effects, which is not a property of the vertex tensor itself, but a property of the external particles, and hence not particularly insightful for this work.
\section{Summary}
In summary, this paper presents a detailed calculation of the one-loop fermionic correction to the effective ALP-photon coupling in the presence of an arbitrary, constant, homogeneous magnetic field. Motivated by the importance of this coupling for ALP detection via the Primakoff effect in astrophysical and terrestrial settings, we aimed to provide an exact result beyond approximations previously considered in the literature.

We employed the Furry picture and utilized fermion propagators dressed by the external magnetic field, derived using Schwinger's proper-time method. The Lorentz structure of the resulting effective vertex tensor, $\Gamma_{a\gamma\gamma}^{\mu\nu\rho}$, was systematically decomposed using the Ritus basis, which naturally incorporates the external field. The core of the work involved the explicit evaluation of the triangle loop diagram. We performed the spacetime and momentum integrations exactly following the methods presented in \cite{PhysRevD.107.116024,Ayala_2024}, carefully handling the Schwinger phase factor arising from the non-perturbative propagators.

The final expression for the vertex tensor involves integrals over three proper-time variables. The integral over the total proper time was addressed using contour integration techniques, yielding a result composed of an integral along the imaginary axis and a sum over residues from poles located on the real axis. While the explicit expressions for these residues are fairly complicated, we outlined how they can be summed and evaluated numerically. An illustrative case with a closed form, considering parallel external particles, was presented to demonstrate the qualitative behavior of the vertex tensor and its dependence on the magnetic field strength. The results were compared with the leading Landau level approximation, showing significant, and non-trivial enhancements in the effective ALP-photon coupling for strong magnetic fields.
The computed vertex tensor provides the necessary theoretical input for refining signal predictions and guiding experimental search strategies. While the exact evaluation is computationally intensive, it provides a complete framework beyond previous approximations. Future investigations could extend this work to include bosonic loop corrections or explore non-trivial external field configurations, further bridging the gap between theory and experiment in the search for ALPs.
\section{Acknowledgements}
I would like to thank my supervisor, Prof. Barbara Jäger, for her invaluable insights and support during our regular meetings. I am also grateful to Jose Medina for kindly answering some of my questions. 
\appendix
\numberwithin{equation}{section}
\section{Field-Dressed Propagators in External Magnetic Fields}\label{app:propagators}

To account for the effects of external background fields, we adopt the Furry picture to decompose the gauge field as
\begin{equation}
    A_\mu = A_\mu^\text{c} + A_\mu^\text{q},
\end{equation}
where $A_\mu^\text{c}$ denotes the classical external field, and $A_\mu^\text{q}$ is the quantized electromagnetic field. The QED Lagrangian then takes the form
\begin{equation}\label{eq:lagrangian_decomposed}
    \mathcal{L}_{\text{EM}} = -\frac{1}{4}F_{\mu\nu}F^{\mu\nu} + \bar{\psi}\left( \underbrace{i\slashed{\partial} - e_f\slashed{A}_\text{c}}_{=: i\slashed{\mathcal{D}}} - m \right)\psi - e\bar{\psi}\slashed{A}_\text{q}\psi,
\end{equation}
where $F^{\mu\nu}$ is the quantized electromagnetic field strength tensor, $\tilde{F}^{\mu\nu}$ its dual, $\psi$ the fermion field, $e_f$ the electromagnetic charge and $m$ the mass of a given fermion. We use slash notation to indicate the contraction of a four-vector with gamma matrices. Treating the quantum component as a perturbation, we absorb the classical part into the fermionic propagator, which we compute non-perturbatively using Schwinger’s proper-time method \cite{PhysRev.82.664}. 

Our goal is to give a general idea for deriving an expression for the Greens function satisfying the equation
\begin{equation}\label{eq:propagator_equation}
    (i\slashed{\mathcal{D}} - m)S_F(x, x') = \delta^4(x-x').
\end{equation}
Unfortunately, simply converting this equation into momentum space is not immediately justifiable, since the gauge field breaks the translation invariance, unless $A_\mu$ is linear in spacetime. To tackle this problem, we choose the Fock-Schwinger gauge,
\begin{equation}
    A_\mu^\text{c}(x) = -\frac{1}{2}F^\text{c}_{\mu\nu}(x^\nu-y^\nu),
\end{equation} 
which restores the translation invariance, when $y=x'$. This point is fixed for each propagator, and therefore for higher order fermionic loops, the translation invariance of the Eq.~\eqref{eq:propagator_equation} only holds on a single interaction vertex. We therefore now generalize this arbitrary gauge, and study the effects of gauge transformations on the Green's function equation. Writing local gauge transformations as 
\begin{equation}
    A_\mu(x) \to A_\mu'(x)=A_\mu(x) + \partial_\mu \alpha(x),
\end{equation}
where $\alpha(x)$ is a scalar function, Eq.~\eqref{eq:propagator_equation} becomes
\begin{align}
    i\delta^4(x-x') &=e^{i\alpha(x)}(i\slashed{\mathcal{D}} - m)e^{-i\alpha(x)}S'_F(x, x') \\
    &= (i\slashed{\mathcal{D}} - m)e^{-i[\alpha(x)-\alpha(x')]}S'_F(x, x').
\end{align}
Simply put, transforming a propagator from a general gauge $A_\mu$ to a gauge $A_\mu^{\text{FS}}$ in the Fock-Schwinger gauge can be done by
\begin{equation}
    S_F(x, x') = e^{i\Phi(x,x')}\ S^{\text{FS}}_F(x, x').
\end{equation}
with
\begin{equation}
    \Phi(x,x^\prime)=-e_f\int\limits_{x^\prime}^x\text{d}\xi^\mu \left[A_{\mu}^\text{c} + \frac{F^\text{c}_{\mu\nu}(\xi-x^\prime)^\nu}{2}\right],
\end{equation}\label{eq:phase_factor}
which is a gauge and spacetime dependent phase factor. The integration variable is contracted with the tensor inside of the integrand. This way, the gauge and translation dependent part is factored out of the propagator, and we can work with the translation invariant part of the propagator, by fixing the point $x^\prime$ suitably. To calculate $S^\text{FS}$, we take the Fourier transform of the equation \eqref{eq:propagator_equation},
\begin{equation}
    S^\text{FS}_F(p) = i(\slashed{p}-e\slashed{A}^\text{FS}-m)^{-1}=(\slashed{p}-e\slashed{A}^\text{FS}+m)\int\limits_0^\infty\text{d}s~e^{is\{\slashed{p}-e_f\slashed{A}^\text{FS}-m^2+i\varepsilon\}}\label{eq:propagator_fourier}
\end{equation}
where we have used the mathematical identity
\begin{equation}
    \frac{1}{C+i\varepsilon} = \int_0^\infty \text{d}s~e^{is(C+i\varepsilon)},
\end{equation}
for $C\in\mathbb{R}$ and $\varepsilon>0$. To bring the integral into a useful form, we multiply Eq.~\eqref{eq:propagator_fourier} with the Dirac operator from both sides, and get
\begin{align}
    1&=\left(\right[\slashed{p}-e_f\slashed{A}^\text{FS}\left]^{2}-m^2\right)\int\limits_0^\infty\text{d}s~e^{is\{\slashed{p}-e_f\slashed{A}^\text{FS}-m^2+i\varepsilon\}}\nonumber\\
    &=\left(p^2+e_f^2\frac{F^{\text{c},\mu\rho}F_\rho^{\text{c},\nu}}{4}\frac{\partial}{\partial p^\mu}\frac{\partial}{\partial p^\nu}-m^2+e_f\frac{F^\text{c}_{\mu\nu}\sigma^{\mu\nu}}{2}\right)\int\limits_0^\infty\text{d}s~e^{is\{\slashed{p}-e_f\slashed{A}^\text{FS}-m^2+i\varepsilon\}}\label{eq:propagator_fourier_2}
\end{align}
where $\sigma^{\mu\nu}=\frac{i}{2}[\gamma^\mu,\gamma^\nu]$. The last step is to solve this equation, which is not very insightful for our purposes. We instead refer the reader to \cite{Hattori_2023} for the details. 
\begin{figure}[!htb]
    \centering
    \includegraphics[width=0.9\textwidth]{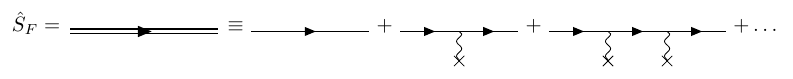}
    \caption{Expansion of the field-dressed fermion propagator in terms of external classical fields.}
    \label{fig:fermion_propagator}
\end{figure}
While the above formulation is general, explicit solutions for the propagator are known only for specific background configurations. In the case of a constant magnetic field, the fermion propagator can be expressed as \cite{PhysRev.82.664,itzykson2006quantum}
\begin{align}
    S_f(X) &= e^{i\Phi(x,x')} \frac{i\beta}{2(4\pi)^2} \int_0^\infty \frac{ds}{s\sin(\beta s)} \Bigg\{ \frac{1}{s} \bigg[ \cos(\beta s) X_\mu \tilde{\Lambda}^{\mu\nu}\gamma_\nu + i\sin(\beta s) X_\mu \hat{\tilde{F}}^{\mu\nu} \gamma_\nu \gamma^5 \bigg] 
    \nonumber \\
    &\quad - \frac{\beta}{\sin(\beta s)} X_\mu \tilde{\Lambda}^{\mu\nu} \gamma_\nu + m_f \big[ 2\cos(\beta s) + \sin(\beta s) \gamma_\mu \hat{F}^{\mu\nu} \gamma_\nu \big] \Bigg\} 
    \nonumber \\
    &\quad \times e^{-i(m^2s + \frac{X_\mu \tilde{\Lambda}^{\mu\nu} X_\nu}{4s} - \frac{\beta}{4\tan(\beta s)} X_\mu \tilde{\Lambda}^{\mu\nu} X_\nu)}.\label{eq:fermion_propagator}
\end{align}
Here, $X^\mu = x^\mu - x^{\prime\mu}$, $\beta = e_f B$, $\Lambda^{\mu\nu} = F^{\mu}_{\text{c},{\mu}\alpha}F_\text{c}^{\alpha\nu} / B^2$, $\tilde{\Lambda}^{\mu\nu} = \tilde{F}^{\mu}_{\text{c},{\mu}\alpha}\tilde{F}_\text{c}^{\alpha\nu} / B^2$, $\hat{F}_\text{c}^{\mu\nu} = F_\text{c}^{\mu\nu} / |B|$, and $\hat{\tilde{F}}^{\mu\nu} = \tilde{F}^{\mu\nu} / |B|$. This propagator can be understood as the resummation of its perturbative expansion in external fields, as illustrated in Figure~\ref{fig:fermion_propagator}. Due to its complexity, the perturbative expansion remains the dominant approach in practical calculations, which is valid for weak magnetic fields ($B \ll B_c$). 
\section{Details on the Strong Field Limit}\label{app:strong_mag}
In this section, we provide the explicit expressions for the integrals appearing in Eq.~\eqref{eq:strong_inter}. We have tensorial integrals of varying ranks, up to rank three, which we have to calculate in $1+1$ dimensions. Typically, similar integrals are calculated in $1+3$ dimensional space-time, where the integrals can be decomposed into sums of smaller scalar integrals. However, most computational tools like \textsc{FeynCalc} are optimized for the case where the space-time dimension is in the vicinity of four, and can therefore spit out erroneous results for our integrals.\\
We therefore calculate the integrals by hand, using the aforementioned assumptions in the main text. By keeping the indices explicit, we identify ten distinctly different integrals, none of which are divergent. To keep the notation simple, we define
\begin{equation}
    I^{ij}=\int \text{d}^2w_\parallel~\frac{(w_{\parallel}^3)^i(w_{\parallel}^0)^j}{[(w_\parallel-k_\parallel)^2-m_f^2+i\varepsilon][(w_\parallel-p_{1\parallel})^2-m_f^2+i\varepsilon][w_\parallel^2-m_f^2+i\varepsilon]}.
\end{equation}
The integrals can then be expressed as
\begingroup
\allowdisplaybreaks
\begin{align}
    I^{00}&=\frac{2i\pi}{m_f^2m_a^2\sqrt{\frac{4m_f^2}{m_a^2}-1}}\tan^{-1}\left(\frac{1}{\sqrt{\frac{4m_f^2}{m_a^2}-1}}\right)\\
    I^{10}&=\frac{i\pi}{m_f^2m_a}\left[1-\sqrt{\frac{4m_f^2}{m_a^2}-1}\tan^{-1}\left(\frac{1}{\sqrt{\frac{4m_f^2}{m_a^2}-1}}\right)\right]\\
    I^{01}&=\frac{i\pi}{m_f^2m_a\sqrt{\frac{4m_f^2}{m_a^2}-1}}\tan^{-1}\left(\frac{1}{\sqrt{\frac{4m_f^2}{m_a^2}-1}}\right)\\
    I^{11}&=\frac{i\pi}{2m_f^2}\left[1-\sqrt{\frac{4m_f^2}{m_a^2}-1}\tan^{-1}\left(\frac{1}{\sqrt{\frac{4m_f^2}{m_a^2}-1}}\right)\right]\\
    I^{20}&=\frac{i\pi}{4m_f^2}\left[3-2\sqrt{\frac{4m_f^2}{m_a^2}-1}\tan^{-1}\left(\frac{1}{\sqrt{\frac{4m_f^2}{m_a^2}-1}}\right)\right]\\
    I^{02}&=-\frac{i\pi}{4m_f^2}\left[1-\frac{2}{\sqrt{\frac{4m_f^2}{m_a^2}-1}}\tan^{-1}\left(\frac{1}{\sqrt{\frac{4m_f^2}{m_a^2}-1}}\right)\right]\\
    I^{30}&=\frac{i\pi}{24m_f^2m_a}\left[6(4m_f^2-m_a^2)\sqrt{\frac{4m_f^2}{m_a^2}-1}\tan^{-1}\left(\frac{1}{\sqrt{\frac{4m_f^2}{m_a^2}-1}}\right)+11m_a^2-24m_f^2\right]\\
    I^{21}&=\frac{i \pi m_a}{8m_f^2}\left[3-2\sqrt{\frac{4m_f^2}{m_a^2}-1}\tan^{-1}\left(\frac{1}{\sqrt{\frac{4m_f^2}{m_a^2}-1}}\right)\right]\\
    I^{12}&=\frac{i \pi m_a}{24m_f^2}\left[5-6\sqrt{\frac{4m_f^2}{m_a^2}-1}\tan^{-1}\left(\frac{1}{\sqrt{\frac{4m_f^2}{m_a^2}-1}}\right)\right]\\
    I^{03}&=-\frac{i \pi}{8m_f^2}\left[3-\frac{2}{\sqrt{\frac{4m_f^2}{m_a^2}-1}}\tan^{-1}\left(\frac{1}{\sqrt{\frac{4m_f^2}{m_a^2}-1}}\right)\right]
\end{align}
\endgroup
As an important note, we would like to point out that after performing a Bose-symmetrization of the vertex tensor, the only difference in these integrals is that the $p_1$ term is replaced by $p_2$. This has the consequence that each integral with an odd number of $w_\parallel^3$ in the numerator changes its sign.
\bibliography{bib}
\bibliographystyle{JHEP}

\end{document}